\documentclass[%
preprint,
 amsmath,amssymb,
]{revtex4}

\usepackage{framed}
\usepackage{graphicx}
\graphicspath{{images/}}

\usepackage{mathrsfs}

\DeclareMathOperator{\arctanh}{arctanh}

\begin{document}

\title{Geometry and proper time of a relativistic quantum clock}

\author{Joseph Balsells}
\email{balsells@psu.edu}
\author{Martin Bojowald}
\email{bojowald@psu.edu}
\affiliation{%
  Institute for Gravitation and the Cosmos,
  The Pennsylvania State University, \\
  104 Davey Lab, 251 Pollock Road,
  University Park, Pennsylvania 16802, USA
}

\begin{abstract}
  Classical clocks measure proper time along their
  worldline, and Riemannian geometry provides tools for predicting the
  time shown by clocks in both flat and curved spacetimes. Common approaches
  to time in quantum systems, based for instance on wave functions or density
  matrices, tend to obscure this geometric property at the quantum
  level. Here, a new framework is demonstrated for perturbing the classical
  path-length functional to include quantum degrees of freedom within a
  modified Riemannian geometry. In this framework, a quantum clock travels on
  geodesics of a family of spacetimes deformed by parameters specifying the
  clock's quantum state. Detailed derivations provide potentially testable
  corrections to gravitational time-dilation in Schwarzschild spacetime that
  scale with the ratio of the clock's Compton wavelength to its wave packet's
  spatial extent.
\end{abstract}


\maketitle

\section{Introduction}
\label{sec:introduction}

General relativity offers a framework for understanding how the classical
degrees of freedom of matter and spacetime relate. However, in its standard
presentation, the apparatus of general relativity is not equipped to address
the quantum degrees of freedom of either matter or spacetime. Quantum field
theory on curved spacetime successfully unifies quantum mechanics and
relativity, but in this case, spacetime merely provides a new stage for
standard quantum concepts such as scattering or particle production. It
remains an open question to determine a systematic way to extend general
relativity and its geometrical picture to encompass quantum degrees of
freedom.

Several quantization schemes have been proposed for quantizing either the
gravitational field, matter, or both simultaneously
\cite{rovelli2008,ali2005,douglas2019}. Although achieving a consistent
simultaneous quantization of both matter and gravitational fields is highly
desirable, the prospects for obtaining direct evidence of such physics are
limited \cite{hossenfelder2013}. An alternative program focuses on addressing
specific points of conflict between quantum theory and relativity. For
example, this has led to foundational studies on the equivalence principle
\cite{zych2018}, the black hole information problem \cite{polchinski2017}, and
the cosmological constant problem \cite{martin2012}. Elucidating any of these
isolated conflicts is hoped to provide insight into a unified theory of
quantum gravity. Another set of issues of this kind is known as the ``problem
of time.''

The problem of time in quantum theory has a number of guises
\cite{KucharTime,anderson2012}. Here, we consider the problem as it arises in
the analysis of the relativistic point particle in curved spacetime. The
relativistic point particle is interesting because it serves as a prototype
for quantizing the gravitational field, which is also governed dynamically by
a constraint \cite{DiracHamGR,Katz,ADM}, as well as for string quantization
\cite{GreenSchwarzWitten2012}. Classically, the time read by an ideal
point-like clock in a curved spacetime is computed as the arc-length of the
clock's worldline. However, if the clock is not point-like but acquires
quantum properties, then it is not immediately clear how to generalize the
notion of proper time read by the clock. For instance, a clock starting in a
superposition of different initial positions would seem to require some kind
of averaged proper times along different classical geodesics. And a clock wave
function spread out in position space should be subject to tidal forces, but
more general ones than a classical extended object due to detailed quantum
features such as correlations, impurity, or entanglement. Addressing these
questions requires a systematic framework that succesfully implements the
geometrical nature of gravity with quantum effects, making sure that an
unambiguous notion of geodesics and proper time still exists. Our new
treatment of the relativistic particle in curved spacetime presents a
promising starting point for this endeavor.

The technical challenges involved in quantizing the relativistic particle are
several. Broadly speaking, the lack of a unique time parameter in the
classical theory is reflected by a conceptually complicated notion of time
evolution in the quantum theory. More technically, imposing spacetime symmetry
on the particle action leads to a reparametrization-invariant Lagrangian, an
example of a gauge theory in which gauge transformations can be used to change
the time coordinate. In this case, the classical evolution of the particle is
constrained to a subspace of the phase space where the Hamiltonian vanishes
\begin{equation}
  \label{eq:58}
  H = 0.
\end{equation}
In canonical quantization, the Hamiltonian constraint selects the physical Hilbert space \(\mathcal{H}_{\text{phys}}\) as the Cauchy completion of states which are annihilated by the constraint operator
\begin{equation}
  \label{eq:21}
  \hat{H}\vert \Psi \rangle = 0.
\end{equation}
However, a consequence of this constraint is that the Schr{\"o}dinger equation
seems to predict no dynamics,
\begin{equation}
  \label{eq:22}
  i\hbar \partial_t \vert \Psi \rangle = \hat{H}\vert \Psi \rangle = 0.
\end{equation}
The system described by the quantum state \(\vert \Psi \rangle\) appears frozen.

One approach to resolving this problem of time may be found in relational
approaches to quantum mechanics, as already suggested in
\cite{dirac1964,BergmannTime}. Relational theories are physically well-motivated: if a
reference frame clock is treated as a physical system with its own degrees of
freedom, it should admit a quantum description. Then, it is possible to
reconsider time not as a global parameter with respect to which all elements
of the system evolve, but as a relational concept in which subsystems may
evolve with respect to each other. A development in this area is the
exploration of quantum reference frames \cite{QFR,GeomObs2,ReferenceFrames,giacomini2019}.

As a simple model, one chooses the multiparticle state consisting of a clock
\(C\) and an observer system \(S\). Conditioning the observer system on the
state of the clock allows changes in the clock state to give a fully
quantum-mechanical realization to time evolution of a quantum system. Choosing
clock degrees of freedom and interactions which correspond to an adequate
notion of time has been pursued under the Page-Wootters formalism
\cite{PageWootters}. Once a clock acquires quantum features, then there is a
possibility of the clock showing a superposition of times. This quantum time
dilation effect is one of the key predictions identified in \cite{smith2020}.

Recent applications of the Page-Wootters formalism depend on novel conceptual
approaches inspired by quantum information theory. This approach seems to
differ greatly from the classical geometric interpretation of time because
spacetime is extended significantly by an infinite-dimensional space of
states. One may ask whether the classical theory might be adapted to
incorporate some (or all) of the features of the quantum Page-Wootters theory,
or at least of a relational quantum state. In this light, in Section
\ref{sec:constr-quant-phase} we adopt a geometric formulation of quantum
theory originally developed for non-relativistic systems in
\cite{strocchi1966,kibble1979,ashtekar1999,bojowald2006}, and generalized to
the relativistic case in \cite{EffCons,EffConsRel}.

In Section \ref{sec:applications}, we apply these phase space methods to the geodesic equation. We find a
canonical phase space description of the quantum relativistic particle in
curved spacetime which retains the same geometrical structure of the classical
metric theory and, when restricted to a Minkowski background, produces quantum
time dilation effects equivalent to those obtained by Smith and Ahmadi in
\cite{smith2020}. In addition to elucidating these questions by providing an
alternative solution, our approach can more easily be generalized to curved
backgrounds. For instance, we apply it to Schwarzschild spacetime and address
the new question of how quantum proper time could modify the approach of a
falling object to a black hole horizon. In particular, our formalism makes it
possible to clarify comparisons between physical observables in different
frames, as discussed in more detail in our final Section~\ref{sec:discussion}.

% Beginning from the classical Lagrangian theory, we develop the corresponding Hamiltonian theory for a relativistic free particle. Coordinate time is initially introduced as an operator on the particle Hilbert space and proper time is used as the evolution parameter. We reformulate the quantum theory in a gauge theory of classical type using the quantum phase space approach developed in \cite{bojowald2006}. Phase space reduction of the resulting quasiclassical theory via gauge-fixing allows us to eliminate any dependence on quantities derived from the time operator. Detailed methods are covered in the Appendix Section \ref{a:detailed}. 

\section{Methods}
\label{sec:constr-quant-phase}

In this section, we develop a geometric framework for describing a quantum relativistic point
particle in curved spacetime. Such an aim is justified provided that the interaction of the quantum field with external backgrounds does not result in particle creation. In that case, the quantum theory we obtain will be equivalent to the one-particle sector of the corresponding quantum field theory \cite{gavrilov2000a}.

We start with the classical theory of a relativistic point particle, defined by a Lagrangian action that makes spacetime symmetry explicit.
This action is the trajectory arclength in an external geometry, called the proper time, and expressed in local coordinates \(x^a\), \(a = 0,1,2,3\), as
\begin{equation}
  \label{eq:classicalProperTime}
  \tau[\gamma] 
  = 
  \int_0^1 \sqrt{-g_{ab}(x) \frac{dx^a}{d\sigma} \frac{dx^b}{d\sigma}} d\sigma.
\end{equation}
Here \(g_{ab}\) is the metric with signature \((-,+,+,+)\) and \(\gamma\colon [0,1]\ni \sigma  \mapsto \gamma(\sigma)\in M\) is a parametrized timelike curve in the spacetime manifold \(M\).

A key feature of classical relativity is that the action for a freely falling
test particle is independent of its properties, embodying the equivalence
principle where the particle's mass \(m\) does not appear in the
equations. However, for non-gravitational interactions, the action must
include a mass factor \(-mc^2\), with the minus sign ensuring the action is
minimized. The quantum perturbations we consider later will typically not
satisfy the weak equivalence principle \cite{balsells2023}. (It is possible to
interpret quantum terms as implying a modified dispersion relation, which is
also known to imply violations of the equivalence principle \cite{WeakEquivDispersion}.) In
anticipation of this, we keep factors of the particle mass even at the
classical level. With these considerations, the action \(S\) for a free
particle is
\begin{equation}
  \label{eq:action}
  S= -mc^2\tau[\gamma],
\end{equation}
where \(\tau[\gamma]\) is the proper time in Eq.~(\ref{eq:classicalProperTime}). 

Our quantization procedure is based on canonical quantization. In
Appendix~\ref{sec:hamilt-form}, we demonstrate that the Lagrangian system is
singular and follow the Dirac-Bergmann algorithm to construct the constrained
Hamiltonian system. The result is the Hamiltonian
\begin{equation}
  \label{eq:classicalProperTimeHamiltonFunction}
  H(x^a,p_a)
  = - \frac{1}{2m}
  \left(
    g^{ab}p_a p_b + (mc)^2
  \right).
\end{equation}
with
\begin{equation}
  p_a=-mg_{ab}\dot{x}^b\,.
\end{equation}
These momenta are canonically conjugate to the position coordinates:
\begin{equation}
  \{x^b,p_a\}=\delta^b_a\,.
\end{equation}

As a remnant of the intermediate gauge formulation, $H(x^a,p_a)$ is a
Hamiltonian constraint that must vanish, $H(x^a,p^a)=0$,
for physical phase-space variables.
Due to this unfixed gauge, the quantum system exhibits a zero-valued
Hamiltonian and time evolution appears trivial.
We can resolve this issue by reformulating the quantum theory into a
quasi-classical phase space theory using the geometrical methods
from \cite{bojowald2006}, as detailed in Appendix~\ref{a:detailed}. Essentially, a formal power series for semiclassical states identifies quantum evolution as perturbative corrections to classical gauge theory evolution.

In the quasiclassical phase space theory, the classical $x^b$ and $p_a$ are interpreted as expectation values of
basic canonical operators in a (possibly mixed) semiclassical state. At the
same time, they are accompanied by additional independent variables, given by
the moments
\begin{eqnarray}
  \Delta \big (x^{\alpha}p^{\beta}\big)&=&\Delta \big ((x^0)^{\alpha_0}\cdots
                                           (x^3)^{\alpha_3}p_0^{\beta_0}\cdots p_3^{\beta_3} \big)\\
  &=& \big\langle
  (\hat{x}^0-x^0)^{(\alpha_0}\cdots(\hat{x}^3-x^3)^{(\alpha_3}(\hat{p}_0-p_0)^{\beta_0)}\cdots(\hat{p}_3-p_3)^{\beta_3)}
  \big\rangle
\end{eqnarray}
using complete symmetrization indicated by parentheses in the superscript
indices, as well as the notation of multi-indices
\(\alpha = (\alpha_0,\alpha_1,\alpha_2, \alpha_3)\) and
\(\beta = (\beta_0, \beta_1,\beta_2,\beta_3)\), such that
\begin{align}
  x^\alpha = (x^0)^{\alpha_0} (x^1)^{\alpha_1} (x^2)^{\alpha_2} (x^3)^{\alpha_3} \\
  p^\beta = (p_0)^{\beta_0} (p_1)^{\beta_1} (p_2)^{\beta_2} (p_3)^{\beta_3}\,.
\end{align}
Any mixed state can therefore be represented by its basic expectation values
and moments used as coordinates on state space, $( x^a, p_a, \Delta(x^\alpha p^\beta))$
for all multi-indices $\alpha$ and $\beta$. 

In future equations, we will use 
further arithmetic rules for the multi-indices, including the factorial
\begin{equation}
  \label{factorial}
  \alpha! = \alpha_0! \alpha_1! \alpha_2 ! \alpha_3!
\end{equation}
and order
\begin{equation}
  \label{order}
  \vert \alpha \vert =
  \alpha_0 + \alpha_1 + \alpha_2 + \alpha_3\,.
\end{equation}
Defining the order of a moment as the value of the sum \(\vert \alpha +
\beta \vert = \sum_i (\alpha_i + \beta_i)\), we call a state semiclassical if
it has moment coordinates such that
\begin{equation}
  \label{eq:hierarchy}
  \Delta(x^\alpha p^\beta) = O(\hbar^{\vert \alpha + \beta \vert /2}).
\end{equation}
This condition is satisfied for Gaussian states, but the set of semiclassical
states defined in this way is much larger because numerical coefficients are
unspecified for all the moments.

Functions on the space of states can be generated by operators $\hat{A}$, such that
$\hat{A}(\psi)$ is defined as the expectation value, ${\rm
  Tr}(\rho_{\psi}\hat{A})$, the operator takes in the state $\psi$ with density
matrix $\rho_{\psi}$. The state space is then equipped with a Poisson bracket
\begin{equation}
  \label{eq:1}
  \left\{ \mathrm{Tr}(\rho_\psi \hat{A}),
    \mathrm{Tr}(\rho_\psi \hat{B})
  \right \}
  = \frac{1}{i\hbar} \mathrm{Tr}(\rho_\psi [\hat{A},\hat{B}])
\end{equation}
extended to moments (which contain products of expectation values) by using
the Leibniz rule. This bracket allows us to compute Hamilton's equations for
any function of basic expectation values and moments. In particular, given a
Hamilton operator $\hat{H}$ that quantizes our Hamiltonian (constraint)
(\ref{eq:classicalProperTimeHamiltonFunction}), using the completely  symmetric choice of
operator ordering if $g^{ab}$ depends on $x^c$, we obtain the quantum
Hamiltonian
\begin{align}
  H_\mathcal{Q}\big (x^a,p_a,\Delta(x^\alpha p^\beta) \big )
  &\equiv \mathrm{Tr}\big(\rho_{\psi} 
    H \big (x^a + (\hat{x}^a - x^a),p_a + (\hat{p}_a - p_a)\big )\big)
    \nonumber \\[1em] 
  &= \sum_{\alpha=0}^\infty\sum_{\beta=0}^\infty
    \frac{1}{\alpha!\beta !} \frac{\partial^{\vert \alpha+\beta \vert }H(x,p)}{\partial^{\alpha}x\partial^{\beta}p} \; \Delta(x^\alpha p^\beta)\,. \label{eq:Hq}
\end{align}

This Hamiltonian generates Hamilton's equations for basic expectation values
and moments, for instance
\begin{equation}
  \frac{d \Delta(x^{\alpha}p^{\beta})}{d\tau} = \{\Delta(x^{\alpha}p^{\beta}),H_{\mathcal{Q}}\}\,.
\end{equation}
It also implies the quantum constraint $H_{\cal{Q}}=0$ as well as higher-order
constraints for the moments. Here, we use results from
\cite{EffCons,EffConsRel} that allow us to solve the constrained problem by
eliminating moments related to the time coordinate $t=x^0$ (which would not be
turned into an operator in non-relativistic quantum mechanics). This step,
which amounts to a choice of gauge fixing in the underlying gauge theory and
is also explained in more detail in Appendix~\ref{a:detailed},
depends on the space-time coordinate system used to express the background
metric $g^{ab}$ as specific functions of the coordinates. In physical
language, this step represents the choice of a reference frame, which does not
affect observables classically because of general covariance. By introducing
moment variables, our methods allow us to determine possible implications of
reference frames in a quantum setting.

\section{Applications}
\label{sec:applications}

We now apply this theory to obtain
corrections to radial geodesics in a static, spherically symmetric spacetime.
The quantum Hamilton function corresponding to the geodesic Hamilton function
in proper time gauge, Eq.~(\ref{eq:classicalProperTimeHamiltonFunction}), is
not easily resummable when expanded in moments. As an approximation to the
full quantum theory, we can truncate the series at second order. Since the
choice of frame eliminates time moments, the only moments left for radial
motion in a background spacetime, which constitutes our main case of interest,
are those involving the radial coordinate $r$ and its momentum, $p_r$. 

With these ingredients, we arrive at the quantum Hamiltonian
\begin{eqnarray}
  \label{eq:Hq2dof}
    H_\mathcal{Q}\big (t,r,p_t,p_r, \Delta(r^2), \Delta(rp_r), \Delta(p_r^2) \big )
    &=& - \frac{1}{2m} \left( g^{ab}p_ap_b + m^2c^2\right) \\
    &&
      - \frac{1}{2m} \Bigg [ 
      \frac{\Delta \left(r^2\right)}{2}
      \left(p_r^2 \partial_r^2 g^{rr} + 2 p_r p_t \partial_r^2 g^{tr} + p_t^2
      \partial_r^2 g^{tt}\right) \nonumber\\
    &&\qquad
      +
      2 \Delta (rp_r)
      \left(p_r \partial_r g^{rr} + p_t \partial_r g^{tr}\right)
      + \Delta \left(p_r^2\right) g^{rr} 
      \Bigg]. \nonumber
\end{eqnarray}
The structure of this expression becomes
clearer when we express the radial moments in terms of canonical coordinates.
The radial second-order moments $\Delta(r^2)$, $\Delta(rp_r)$ and
$\Delta(p_r^2)$ have Poisson brackets
\begin{equation}
  \{\Delta(r^2),\Delta(p^2)\}=4\Delta(rp_r)\quad,\quad
  \{\Delta(r^2),\Delta(rp_r)\}=2\Delta(r^2)\quad,\quad
  \{\Delta(rp_r),\Delta(p_r^2)\}=2\Delta(p_r^2)
\end{equation}
which are not canonical. A non-linear change of variables given by
\begin{equation}
  \label{eq:canonicalCoordinates}
  \Delta \left(r^2\right) = s^2
  \qquad
  \Delta (p_r r) = s p_s
  \qquad
  \Delta \left(p_r^2\right) = p_s^2 + \frac{U}{s^2}
\end{equation}
results in a canonical pair $\{s,p_s\}=1$ and a so-called Casimir variable
$U$ that has vanishing Poisson brackets with both $s$ and $p_s$.  (This
transformation, without the background of Poisson geometry, has been found
several times independently in a variety of fields
\cite{VariationalEffAc,GaussianDyn,QHDTunneling}. A derivation from Poisson
geometry and generalizations to higher orders and two degrees of freedom can
be found in \cite{Bosonize,EffPotRealize}.)  Its value equals the uncertainty
product, $U=\Delta(r^2)\Delta(p_r^2)-\Delta(rp_r)^2$ and is therefore bounded
from below by $\hbar^2/4$. Higher order relations are developed for quantum
states in \cite{brizuela2014}, although we will not need them here.  The
variable $U$ can be related to a second canonical pair by defining
\begin{equation}
  U = p_q^2 \frac{\hbar^2}{4}
\end{equation}
with the momentum $p_q$ of a canonically conjugate variable $q$ that does not
appear elsewhere in the Hamiltonian (implying that $p_q$, just like $U$, is
conserved). With this choice, momentum fluctuations $\Delta(p_r^2)$ are
homogeneous of degree two in the canonical momenta, $p_s$ and $p_q$. In the
quantum theory \(\vert p_q \vert \ge 1\), and the classical limit corresponds
to \(\vert p_q \vert \rightarrow 0\).

The set of all momenta is now given by
\(p_{\bar{a}} = (p_t,p_r,p_s,p_q)\). The barred index emphasizes that quantum
degrees of freedom have extended the original, classical index set.  With
these considerations, the quantum Hamilton function can be expressed concisely
as
\begin{equation}
  \label{eq:HQgeodesic}
  H_{\mathcal{Q}}(t,r,s,q,p_t,p_r,p_s,p_q) = - \frac{1}{2m} \left(
    g_{\mathcal{Q}}^{\bar{a}\bar{b}} p_{\bar{a}}p_{\bar{b}} + m^2c^2\right). 
\end{equation}
The symmetric matrix associated to the quadratic form is a deformation of the
classical metric and equals 
\begin{equation} \label{eq:gQinv}
  g_{\mathcal{Q}}^{\bar{a}\bar{b}} :=
  \begin{pmatrix}
    g^{tt} + \frac{1}{2}s^2 \partial_r^2g^{tt}
    & g^{tr} + \frac{1}{2}s^2 \partial_r^2g^{tr} & s \partial_r g^{tr} & 0
    \\
    g^{tr} + \frac{1}{2}s^2 \partial_r^2g^{tr}
    & g^{rr} + \frac{1}{2}s^2 \partial_r^2 g^{rr} & s\partial_r g^{rr} & 0
    \\
    s \partial_r g^{tr} & s\partial_r g^{rr} & g^{rr} & 0
    \\
    0 & 0 & 0 & \frac{1}{4}\hbar^2 g^{rr}/s^2
  \end{pmatrix}.
\end{equation}

\subsection{Quantum proper time}
\label{sec:quantum-proper-time}

An inverse Legendre transform of the quantum Hamilton function,
Eq.~(\ref{eq:HQgeodesic}) yields the quantum Lagrange function.  The transform
is well-defined because the quantum Hamilton function is quadratic in the
momenta and, hence, non-singular. The inverse transform uses the velocities
\((\dot{t}, \dot{r},\dot{s}, \dot{q})\) which are defined through appropriate
partial derivatives of the quantum Hamilton function.

We can organize the quantum Lagrange function by defining the
quantum-effective metric as the matrix inverse of a quadratic form, Eq.~(\ref{eq:gQinv}):
\begin{equation}
  \label{eq:gQ}
  g^{\mathcal{Q}}_{\bar{a}\bar{b}} := \left(g_{\mathcal{Q}}^{\bar{a}\bar{b}}\right)^{-1}.
\end{equation}
Direct computation reveals that
\begin{equation}
  \label{eq:LQ}
  L_{\mathcal{Q}} = -\frac{1}{2} m g^{\mathcal{Q}}_{\bar{a}\bar{b}} \dot{x}^{\bar{a}} \dot{x}^{\bar{b}} + \frac{m c^2}{2}.
\end{equation}
For small moments in the semiclassical approximation, the quantum metric can
be inverted perturbatively, such that
\begin{equation} \label{eq:gsemiclass}
  g_{\bar{a}\bar{b}}^{\mathcal{Q}}\dot{x}^{\bar{a}}\dot{x}^{\bar{b}}=g_{ab}\dot{x}^a\dot{x}^b+O(\Delta^2(\cdot))
\end{equation}
where $\Delta^2(\cdot)$ stands for any expression of homogeneity degree two in the
quantum variables, $s$, $\dot{s}$, $\dot{q}$ as well as $\sqrt{\hbar}$.

When the Hamilton function satisfies the constraint \(H_{\mathcal{Q}} = 0\), the Lagrange function satisfies the corresponding constraint
\begin{equation}
  \label{eq:LagrangianConstraint}
  L_{\mathcal{Q}} = mc^2.
\end{equation}
The Lagrangian equations of motion of \(L_{\mathcal{Q}}\), subject to the
constraint~(\ref{eq:LagrangianConstraint}), are equivalent to Hamilton's
equations for the constrained system governed by \(H_{\mathcal{Q}}\). 

The Lagrangian equations can be obtained as the stationary solutions to a reparametrization-invariant quantum proper time path functional 
\begin{equation}
  \label{eq:33}
  \tau_{\mathcal{Q}} := 
  \int \sqrt{-g_{\bar{a}\bar{b}}^{\mathcal{Q}} dx^{\bar{a}} dx^{\bar{b}}}.
\end{equation}
Here, the quantum-effective metric defines an effective line element
\begin{equation}
  \label{eq:dsQ}
  ds^2 =
  g_{\bar{a}\bar{b}}^{\mathcal{Q}}(r,s) dx^{\bar{a}}dx^{\bar{b}}
\end{equation}
for the purpose of defining the line element. (It is interesting to ask
whether this quantum effective metric obeys an extended covariance condition
with a higher-dimensional tensor transformation law that leaves the effective
line element invariant. We leave this as an open question, but note that
recent work on quantum reference frames, as mentioned in our Discussion
section \ref{sec:discussion} suggests that this task may be challenging.)

On the part of the quantum phase space where quantum states obey the semiclassical hierarchy condition, Eq.~(\ref{eq:hierarchy}), the quantum Lagrange function splits into classical and non-classical contributions. In such cases, the quantum proper time obtains dominant contributions from motion through the classical geometry with sub-dominant contributions from quantum effects:
\begin{equation}
  \label{eq:tauQ}
  \begin{aligned}
    \tau_{\mathcal{Q}}
    &= \int L_{\mathcal{Q}} d\sigma \\
    &= \int \left( L_{\text{classical}} + L_{\text{non-classical}} \right) d\sigma \\
    &= \tau_{\text{classical}} + \tau_{\text{non-classical}}.
  \end{aligned}
\end{equation}
The additive nature of this expression is implied by the general semiclassical
result (\ref{eq:gsemiclass}) for the metric.

As an example, we compute the proper time for a semiclassical quantum state
in-falling from a finite position in Schwarzschild spacetime.
For an ideal point-like clock which was initially at rest at \(r_0\), the classical part of the proper time read at some \(r<r_0\) is a standard result,
\begin{equation}
  \label{eq:14}
  \tau_{\text{classical}}(r)
  =
  \frac{r_0}{c}
  \left[
    \sqrt{\frac{r_0}{r_s}} \arctan\left(\sqrt{\frac{r_0}{r}-1}\right)
    +
    \frac{r}{r_s} \sqrt{\frac{r_s}{r}-\frac{r_s}{r_0}}
  \right]
\end{equation}
with the Schwarzschild radius $r_s$.
If the clock is modeled not as an ideal point object, but has quantum features, then our theory permits us to compute leading-order corrections to the classical proper time. In the quantum-adiabatic limit, \(\dot{s}=0\), these are obtainable in closed-form as
\begin{equation}
  \label{eq:tauNonClassical}
  \begin{split}
    \tau_{\text{non-classical}}(r) 
    &=  \left(\frac{p_q \hbar}{2 m c s}\right)^2
      \left(
      \left( 1
      - \frac{3}{2} \frac{r_s}{r_0} \right) \tau_{\text{classical}}
      + \frac{r}{c}  \sqrt{   \frac{r_s}{r} - \frac{r_s}{r_0}   }
      \right) \\
    &
      - \frac{1}{2} \frac{r_s}{c}
    \frac{s^2}{r_0^2}
      \left(
    \frac{1}
    {\sqrt{r_s/r - r_s/r_0}}
      +
      3
    \frac{
    \sqrt{r_0/r} \arctan\left(\sqrt{r_0/r-1}\right)}
    { 1-r_s/r_0 }\right.\\
&\qquad\qquad\qquad\qquad\qquad\qquad\left.    +
    \frac{2\arctanh\left(\sqrt{\frac{r_0/r-1}{r_0/r_s-1}}\right)}
    {\left(1-r_s/r_0\right)^{3/2}}
    \right)
  \end{split}
\end{equation}
The terms in this expression are determined by partial derivatives of the
classical metric that appear in (\ref{eq:gQinv}), evaluated at the classical
$r(\sigma)$ and then integrated over $\sigma$. In this quantum adiabatic
approximation, $s$ as well as $p_q$ are determined by moments of the initial state.

This result highlights the general geometrical nature of the separation of
proper time into classical and quantum contributions, found earlier for
specific wave functions in \cite{smith2020}, and it generalizes this property to any
semiclassical state. Our parameterization in terms of $p_q$ and $s$ further
shows that there are two separate contributions to the quantum term since
leading order corrections scale with two competing effects. One, an effect
that originates in the $U$-term and enforces the uncertainty principle (see
\cite{Description}), is proportional to $p_q^2$ as well as the squared Compton
wavelength of the state and tends to increase the proper time read by the
clock. The second is a spatial delocalization effect proportional to the
state's position
variance $s^2$ and tends to decrease the time read by the clock. Notice the
mass dependence of the first term, which is a sign of quantum violations of
the equivalence principle as in \cite{balsells2023}.

This quantum effect on proper time of a clock depends on geometric properties
of a single clock's (extended) worldline. For infinitesimal trajectories, one
clock's proper time can be related to the time read by any other clock by
various time dilation formulas. We obtain these in our next discussion.

\subsection{Time dilation}
\label{sec:time-dilation}

The absence of a preferred time coordinate is a feature of general relativity, highlighting the theory's principle that the measurement of time is relative to the observer's frame of reference.
Consequently, in the classical theory, two clocks, \(A\) and \(B\), measure proper times \(\tau_A\) and \(\tau_B\) which disagree depending on their relative position and motion (but independently of other properties such as their masses or compositions).

We quantify the first-order difference in the rates of two clocks by a Lorentz factor
\begin{equation}
  \label{eq:relativeTimeAB}
  \frac{d\tau_A}{d\tau_B} = \gamma (x_A,x_B,v_{A},v_{B}).
\end{equation}
To write a coordinate expression giving the functional form of \(\gamma\), it is necessary to choose a slicing of spacetime. Time dilation formulas therefore require three choices: the worldline of each of the two clocks and also a choice of coordinate slicing.

In relativity, it is common to align the slicing of spacetime with the reference frame of one of the two clocks, reducing the complexity of the problem.
This choice introduces an asymmetry: the motion and position of the clocks are no longer on equal footing. One clock becomes anchored in the coordinate system, while the other is defined relative to it. This asymmetry implies that, in general, there is no straightforward relationship between the coordinate expressions for \(d\tau_A/d\tau_B\) and \(d\tau_B/d\tau_A\).

By treating one of the clocks implicitly as the coordinate observer (or a
global frame), the expression (\ref{eq:relativeTimeAB}) represents the 0-component of the tangent 4-vector to the worldline of the other clock. If the second clock is freely-falling, then its tangent vector components are determined by the geodesic equation. 
In the classical theory, the geodesic Hamilton function, Eq.~(\ref{eq:classicalProperTimeHamiltonFunction}), produces two equations for the tangent vector components of radial geodesics:
\begin{equation}
  \label{eq:tdotPhaseSpace}
  \dot{t} = c
  \sqrt{
    \vert\mathrm{det}(g^{ab})\vert \frac{p_r^2}{m^2 c^2} -  g^{tt}
  }
\end{equation}
\begin{equation}
  \label{eq:rdotPhaseSpace}
  \dot{r} = \frac{
    \vert\mathrm{det}(g^{ab})\vert p_r
    + g^{tr}
    \sqrt{
      \vert\mathrm{det}(g^{ab})\vert p_r^2 - m^2 c^2 g^{tt}
    }}
  {m g^{tt}}.
\end{equation}

Equation~(\ref{eq:tdotPhaseSpace}) gives the time dilation factor as a phase space function. We can instead write it as a function on the tangent
bundle (using velocity components) by solving the equation
(\ref{eq:rdotPhaseSpace}) for \(p_r\) and substituting. The solution for the
classical time dilation factor obtained in this way has a surprising form
which depends on the deparametrized velocity of the non-coordinate clock, denoted in
two spacetime dimensions as
\begin{equation}
  \label{eq:17}
  \frac{\dot{x}^a}{\dot{t}} = \frac{dx^a}{dt} =  \left( c, \frac{dr}{dt} \right).
\end{equation}
With this definition, solving the Hamiltonian geodesic equation yields
\begin{equation}
  \label{eq:classicalGeneralTimeDilation}
  \gamma_{\text{classical}} = \frac{1}{\sqrt{-g_{ab} \frac{dx^a}{dt} \frac{dx^b}{dt}/c^2 }}.
\end{equation}

In the quantum-perturbed case, Hamilton's equations involve two additional tangent vector components:
\begin{align}
  \label{eq:sdot}
  \dot{s} = \frac{\partial H_{\mathcal{Q}}}{\partial p_s} \\
  \dot{q} = \frac{\partial H_{\mathcal{Q}}}{\partial p_q}\,.
\end{align}
Still, an analogous result to (\ref{eq:classicalGeneralTimeDilation}) holds in the quantum theory if we make the substitution \(g_{ab} \rightarrow g^{\mathcal{Q}}_{\bar{a}\bar{b}}\) and use the deparametrized tangent-vector velocities
\begin{equation}
  \label{eq:17}
  \frac{\dot{x}^{\bar{a}}}{\dot{t}} = \frac{dx^{\bar{a}}}{dt} =  \left( c, \frac{dr}{dt}, \frac{ds}{dt}, \frac{dq}{dt} \right) .
\end{equation}
Indeed, the form of the result, Eq.~(\ref{eq:classicalGeneralTimeDilation}), is determined directly from the constraint
\begin{equation}
  \label{eq:9}
  L = -\frac{m}{2}g_{ab}\dot{x}^a\dot{x}^b + \frac{mc^2}{2} = mc^2 .
\end{equation}
The quantum dynamics are determined by a Lagrangian with the same form obeying the same constraint. Dividing both sides of the constraint by \(\dot{t}\) and rearranging yields immediately the deparametrized time dilation formula
\begin{equation}
  \label{eq:24}
  \frac{dt}{d\tau} \equiv \gamma = \frac{1}{\sqrt{-g^{\mathcal{Q}}_{\bar{a}\bar{b}} \frac{dx^{\bar{a}}}{dt} \frac{dx^{\bar{b}}}{dt}/c^2 }}.
\end{equation}

In the quantum-adiabatic limit when \(\dot{s}=0\) and neglecting the classical Doppler effect by setting \(\dot{r}=0\) produces the special case
\begin{equation}
  \label{eq:10}
  \gamma = \gamma_{\text{classical}} + \gamma_{\text{non-classical}}
\end{equation}
where \(\gamma_{\text{classical}}\) is the classical factor, Eq.~(\ref{eq:classicalGeneralTimeDilation}), and the non-classical contribution is
\begin{equation}
  \label{eq:gammaNC}
  \gamma_{\text{non-classical}} =
  \gamma_{\text{classical}}^3 \left(\frac{g_{tt}g_{tt}'' - 2 (g_{tt}')^2}{g_{tt}}\right) \frac{s^2}{4}
  +
  \gamma_{\text{classical}}^3 \left( \frac{\hbar}{2mc s}\right)^2 \frac{p_q^2}{2}\frac{g_{tt}}{\det{g_{ab}}}
\end{equation}

Using the standard Schwarzschild slicing in which the line element takes the form:
\begin{equation}
  \label{eq:slicingSD}
  g_{ab}dx^adx^b = - \left( 1 - \frac{r_s}{r} \right) c^2 dt^2
  + \frac{1}{1 - r_s/r} dr^2
\end{equation}
gives
\begin{equation}
  \gamma_{\text{classical}} = \frac{1}{\sqrt{1 - r_s/r_0}}
\end{equation}
and
\begin{equation} \label{eq:gammaNCSchwarzschild}
  \gamma_{\text{non-classical}}
  =
  \frac{1}{2}\frac{s^2}{r^2} \frac{r_s/r}{\left(1 - r_s/r\right)^{5/2} }
  +
  \frac{p_q^2}{2} \left(\frac{ \hbar }{2mc s} \right)^2  \left(1 - \frac{r_s}{r} \right)^{1/2}\,.
\end{equation}

The time dilation factor has quantum corrections which scale in the same way as the proper time corrections, Eq.~(\ref{eq:tauNonClassical}). The first correction predicts that in a curved spacetime a poorly localized clock experience greater dilation than a clock in a more localized state. This tidal effect can be realized classically by a finite-size oscillator clock. The second, proportional to \(\hbar^2\), is strictly quantum mechanical and expected even in flat spacetime. It scales according to the ratio of the particle's Compton wavelength to the spatial extent of its wavefunction, as noted independently in \cite{smith2020}.

The reference \cite{smith2020} obtained this quantum time-dilation effect from
the Page-Wootters formalism applied to two clocks, A and B, in a superposition
of momentum states. In \cite{smith2020} the dilation effect appears when one
considers the probability that the first clock reads time \(\tau_A\)
conditioned on the clock B reading time \(\tau_B\). Our formalism produces a
similar result: the coordinate \(\tau\) measures the proper time of an
in-falling clock, while \(t\) represents the proper
time of the coordinate clock.

Time dilation depends on the choice of clocks, and nominally the Painleve-Gullstrand slicing in which the line element takes the form
\begin{equation}
  \label{eq:slicingPG}
  g_{ab}dx^adx^b = - \left( 1 - \frac{r_s}{r} \right) c^2 dt^2
  + 2 c \sqrt{\frac{r_s}{r}} dt dr
  + dr^2.
\end{equation}
implies a separate class of coordinate clocks. However, (\ref{eq:slicingPG}) gives the same form for the 00-component of the metric. Therefore the time-dilation prediction in the adiabatic, static case \(\dot{r} = \dot{s} = 0\), given by (\ref{eq:gammaNC}) is identical to the Schwarzschild-slicing prediction.

The two slicings differ most near the classical horizon, where the metric functions exhibit significantly distinct behaviors. This suggests that non-classical reference frame effects between these two cases will be most apparent in that regime.

% For numbers we can consider the experiment of Kovachy et al.\ \cite{kovachy2015} in which quantum superposition is obtained at the half-meter scale.

\subsection{Signature change}
\label{sec:signature-change}

The spectrum of the metric tensor is denoted \(\lambda(g_{ab}) = \{\lambda_0, \ldots, \lambda_n\}\). Proper orthochronous basis changes can rescale the eigenvalues but not alter their signs. Therefore, the metric signature, defined by the signs of the eigenvalues, is invariant. 
In general relativity, the metric tensor has a Lorentzian signature, represented in two spacetime dimensions by the tuple \((-,+)\). Such a Lorentzian-signature metric distinguishes between time-like, space-like, and light-like intervals.

The result (\ref{eq:gQinv}) for the quantum effective metric predicts that
quantum effects will shift the locations of the zeros and singularities of the
metric functions. This has implications for the causal structure of spacetime
experienced by a quantum object.  We demonstrate these notions in the standard
slicing of the Schwarzschild metric where the eigenvalues of the inverse
metric are:
\begin{gather}
  \label{eq:classicalSchwarzschildEigenvalue0}
  \lambda_t = - \frac{1}{ 1-r_s/r }, \\
  \label{eq:classicalSchwarzschildEigenvalue1}
  \lambda_r = 1-r_s/r.
\end{gather}

For \(r>r_s\), the eigenvalue corresponding to the first eigenvector is negative, while the eigenvalue for the second eigenvector is positive. Although the standard slicing leads to poorly defined coordinates at the Schwarzschild radius, continuing the coordinates beyond this point results in a flip in the signs of the eigenvalues. Because the eigenvalues flip signs at the same point, the signature of the metric is preserved, while the causal nature of the basis vectors is reversed.

Computing eigenvalues for the quantum-effective inverse metric in the standard slicing yields
\begin{gather}
  \label{eq:32}
  \lambda_t^{\mathcal{Q}} = - \frac{1}{ 1-r_s/r } - \frac{r_s/r}{\left( 1 - r_s/r \right)^3} \frac{s^2}{r^2}, \\
  \lambda_{r-s}^{\mathcal{Q}} = 1-\frac{r_s}{r} - \frac{r_ss}{r^2} - \frac{r_ss^2}{2r^3},\\
  \lambda_{r+s}^{\mathcal{Q}} = 1-\frac{r_s}{r} + \frac{r_ss}{r^2} - \frac{r_ss^2}{2r^3}, \\
  \lambda_q = \frac{\hbar^2}{4s^2} \left( 1-\frac{r_s}{r} \right).
\end{gather}
Positivity of the eigenvalue \(\lambda_q\) indicates that the quantum coordinate \(q\) is spacelike for \(r>r_s\). However, due to the block diagonal form of \(g_{\mathcal{Q}}^{\bar{a}\bar{b}}\), it plays no further role in the current discussion.

The remaining three eigenvalues collapse to the classical ones,
(\ref{eq:classicalSchwarzschildEigenvalue0}) and
(\ref{eq:classicalSchwarzschildEigenvalue1}), in the classical limit. For any
$s$, the sign
of the first eigenvalue flips from negative to positive at \(r= r_s\), as in
the classical case. The new quantum coordinate \(s\) splits the classical
eigenvalue (\ref{eq:classicalSchwarzschildEigenvalue1}) into two distinct values,
\(\lambda_{r-s}^{\mathcal{Q}}\) and
\(\lambda_{r+s}^{\mathcal{Q}}\), which change signs at \(r \approx r_s + s\) and
\(r \approx r_s - s\), respectively. Table \ref{tab:schwarzschildSignature} summarizes the
global structure of signature change.

\begin{table}
  \begin{tabular}{c | c c c c}
    & I & II & III & IV\\
    \hline
    \(\lambda_t^{\mathcal{Q}}\) & \(+\) & \(+\) & \(-\) & \(-\)  \\
    \(\lambda_{r-s}^{\mathcal{Q}}\) & \(-\) & \(-\) & \(-\) & \(+\)  \\
    \(\lambda_{r+s}^{\mathcal{Q}}\) & \(-\) & \(+\) & \(+\) & \(+\)  \\
    \(\lambda_q^{\mathcal{Q}}\) & \(-\) & \(-\) & \(+\) & \(+\)  \\
  \end{tabular}
%  \caption{Eigenvalue signs of the inverse quantum metric \(g_{\mathcal{Q}}^{\bar{a}\bar{b}}\) computed in the standard slicing of Schwarzschild. The signature changes across four regions: Region I corresponds to \(r \in (0, r_s - s)\), Region II to \(r \in (r_s - s, r_s)\), Region III to \(r \in (r_s, r_s + s)\), and Region IV to \(r \in (r_s + s, \infty)\).}
    \caption{Eigenvalue signs of the inverse quantum metric \(g_{\mathcal{Q}}^{\bar{a}\bar{b}}\) in the standard slicing of Schwarzschild metric, across four regions: Region I \(\left(r \in (0, r_s - s)\right)\), Region II \((r \in (r_s - s, r_s))\), Region III \((r \in (r_s, r_s + s))\), and Region IV \((r \in (r_s + s, \infty))\).}
  \label{tab:schwarzschildSignature}
\end{table}

For states within their correlation distance of the classical horizon, the effective metric governing their geodesic motion has two time-like directions.
Precise causal implications for states both far from the horizon and near the horizon can be captured by the local light-cone structure. We compute this light-cone structure in Section \ref{sec:lightcone-slope}.

Before then we address the matter of signature change when the quantization is computed in a slicing adapted to the region near the horizon. The Painleve-Gullstrand slicing of the Schwarschild metric, Eq. (\ref{eq:slicingPG}),  produces the eigenvalues 
\begin{gather}
  \label{eq:classicalPGEigenvalue0}
  \lambda_t = - \frac{1}{2} \frac{r_s}{r} \left( \sqrt{ 1 + 4\frac{r^2}{r_s^2} } + 1 \right), \\
  \label{eq:classicalPGEigenvalue1}
  \lambda_r = + \frac{1}{2} \frac{r_s}{r} \left( \sqrt{ 1 + 4\frac{r^2}{r_s^2} } - 1 \right).
\end{gather}
The signs of these eigenvalues remain unchanged across the classical horizon, with (\ref{eq:classicalPGEigenvalue0}) staying timelike and (\ref{eq:classicalPGEigenvalue1}) staying spacelike.

Quantization in these coordinates leads to an effective metric with eigenvalues
\begin{gather}
  \label{eq:32}
  \lambda_{t}^{\mathcal{Q}} = \lambda_t
  + \left(
    \frac{\lambda_t ( \lambda_t -1)}{\lambda_r + (r/r_s) (\lambda_t - \lambda_r -2)}
  \right) \frac{s^2}{r^2}
  ,\\
  \lambda_{r}^{\mathcal{Q}} = \lambda_r
  + \left(
    \frac{\lambda_r ( \lambda_r -1)}{\lambda_t + (r/r_s) (\lambda_r - \lambda_t -2)}
  \right) \frac{s^2}{r^2}
  , \\
  \lambda_0^{\mathcal{Q}} = 1 - \frac{r_s}{r} - \frac{r_s}{r} \left( 1 - \frac{r_s}{r}\right) \frac{s^2}{r^2}, \\
  \lambda_q = \frac{\hbar^2}{4s^2} \left( 1-\frac{r_s}{r} \right)
\end{gather}
with \(\lambda_t\) and \(\lambda_r\) in these expressions given by (\ref{eq:classicalPGEigenvalue0}) and (\ref{eq:classicalPGEigenvalue1}).
Corrections to the classical eigenvalues computed in this slicing do not cause
sign changes in \(\lambda_{t}^{\mathcal{Q}}\) and
\(\lambda_{r}^{\mathcal{Q}}\). The horizon structure therefore appears
different in the two slicings considered here.

\subsection{Volume element}
\label{sec:volume-element}

The determinant of the metric, computed as the product of the eigenvalues, is a coordinate-dependent construction which plays an important role in the geometric theory.
The volume element in curved spacetime is given in terms of the metric determinant as
\begin{equation}
  \label{eq:25}
  dV = \sqrt{- \det{g_{ab}} } \, d^4x
\end{equation}
where the dimension four refers to the four classical spacetime coordinates,
integrating over $t$, $r$, $\vartheta$ and $\varphi$ in a spherically
symmetric geometry.

The determinant of the inverse quantum metric (\ref{eq:gQinv}) extends the classical metric determinant. Computing directly we find
\begin{equation}
  \label{eq:Detginverse}
  \begin{split}
    \det\left(g^{\bar{a}\bar{b}}_\mathcal{Q}\right)
    = \Bigg[
    \det{(g^{ab})}
    &+ \frac{s^2}{2}
    \left(
      g^{rr} \det{(g^{ab})}'' - 2 {g^{rr}}' \det{(g^{ab})}'
    \right) \\
    & +   \frac{s^4}{4}
  \Big( g^{rr} \det{( {g^{ab}}'')}
    4 {g^{rr}}'  {g^{tr}}'  {g^{tr}} ''
      \\
    &\qquad - 2 ( {g^{tr}}')^2  {g^{rr}}'' - 2 ( {g^{rr}}')^2  {g^{tt}} ''
  \Big)
    \Bigg]
    \cdot
    \frac{ \hbar^2 g^{rr} }{ 4s^2}.
  \end{split}
\end{equation}
This expression introduces series corrections associated with quantum
fluctuations and an overall scaling associated with the quantum uncertainty
coordinate \(q\).

The inverse of this determinant provides the volume element
in four dimensions but only for radial motion, integrating over $t$, $r$, $s$
and $q$. This manifold may be extended to include angular coordinates
$\vartheta$ and $\varphi$ as well as their canonical moment coordinates
analogous to $s$ and $q$ for each degree of freedom. If all these values are
included, we obtain a 10-dimensional volume element.

% The series corrections appear as fourth-order terms, with lower-order corrections vanishing when the determinant of the metric is constant (\(\partial_r \det{(g^{ab})} = 0\)). While the fourth-order terms are not fully captured by a second order truncation, after computing the square root, the quantum fluctuations of the volume element are properly accounted for at the correct order.

% In the quantum phase space, the quantum state is viewed as a point. However, viewed on the classical spacetime, the state resembles a fluid. In a second order truncation, the uncertainty product \(U\) is conserved exactly. The fluid is therefore subjected to a conservation law which extends the classical conservation law
% \begin{equation}
%   \label{eq:conservation}
%   \nabla_a T^{ab} = 0.
% \end{equation}
% The overall factor scaling the metric determinant in Eq. (\ref{eq:Detginverse}), ensures the generalized form of (\ref{eq:conservation}) maintains the correct divergence structure, aligning the flow of quantum uncertainty with the geometry of spacetime.

\subsection{Lightcone slope}
\label{sec:lightcone-slope}

Causal structure is locally determined by the slope of light cones. Material objects, including clocks, follow time-like worldlines whose tangent vectors always remain within the lightcone at each point, indicating that they experience the passage of time. In Section \ref{sec:time-dilation}, we argued that a freely falling clock's tangent vector is altered by its quantum state, and in Section \ref{sec:signature-change}, we demonstrated that quantum effects modify the lightcone structure. Here, we check if these predictions are consistent,leading to a generalized, state-dependent causal structure.

In the classical theory, the tangent vector components to a radial null curve, \(V^a = (\dot{t}, \dot{r})\), satisfy
\begin{equation}
  \label{eq:19}
  g_{ab}V^a V^b = 0.
\end{equation}
This quadratic equation can be solved for the lightcone slope
\begin{equation}
  \label{eq:29}
  \left( \frac{dt}{dr} \right)_{\text{lightcone}}
  := \frac{ \dot{t} }{ \dot{r} }
  = \frac{-g_{tr} \pm \sqrt{-\det{(g_{ab})}}}{g_{tt}}.
\end{equation}
In the Schwarzschild slicing, Eq.~(\ref{eq:slicingSD}), the lightcone slope of in-falling and out-going lightrays is given by
\begin{equation}
  \left( \frac{dt}{dr} \right)_{\text{lightcone}}
  = \pm \frac{1}{c} \frac{1}{1 - r_s/r}.
\end{equation}

In the same coordinates, a classical point particle in-falling from finite position \(r_0\) has tangent vector components
\begin{equation}
  \label{eq:SDtangentVector}
  \begin{aligned}
    \frac{dt}{d\tau} &= \frac{1}{1 - r_s/r} \sqrt{ 1 - \frac{r_s}{r_0} } \\
    \frac{dr}{d\tau} &= - c \sqrt{ \frac{r_s}{r} - \frac{r_s}{r_0} }.
  \end{aligned}
\end{equation}
These tangent vector components determine the slope of the massive particle's worldline:
\begin{equation}
  \left( \frac{dt}{dr} \right)_{\text{massive particle}} = \frac{ dt/d\tau }{ dr/d\tau }
  = - \frac{1}{c} \frac{1}{1 - r_s/r} \sqrt{\frac{r}{r_s}} \sqrt{\frac{1 - r_s/r_0}{1 - r/r_0}}.
\end{equation}
When \(r>r_s\) the trajectory of a massive particle remains always within the lightcone:
\begin{equation}
  \label{eq:classicalLightConeCondition}
  \left\vert
    \frac{\left( dt/dr \right)_{\text{massive particle}}}{\left( dt/dr \right)_{\text{lightcone}}}
  \right\vert
  = \sqrt{\frac{r}{r_s}} \sqrt{\frac{1 - r_s/r_0}{1 - r/r_0}} > 1.
\end{equation}

In the quantum-perturbed case, the particle's tangent vector components are modified by tidal and quantum uncertainty effects. For a massive clock with quantum fluctuation \(s = \sqrt{\Delta(r^2)}\) freely falling from finite position \(r_0\), the tangent vector components now satisfy
\begin{equation} \label{eq:quantumTangentVector}
  \frac{dt}{d\tau}
  =
  \frac{\sqrt{1 - r_s/r_0}}{1 - r_s/r }
  \frac{1 + \frac{s^2}{r^2} \frac{r_s/r}{\left(1 - r_s/r\right)^2 }}
  {\sqrt{1 + \frac{s^2}{r_0^2} \frac{r_s/r_0}{\left(1 - r_s/r_0\right)^2 }}}
  \left(
    1
    +
    \left(\frac{ p_q \hbar }{2mcs} \right)^2  \left(1 - \frac{r_s}{r_0} \right)
  \right)^{1/2}
\end{equation}
and
\begin{equation}
  \begin{split}
  \frac{dr}{d\tau}
    =
    &- c \sqrt{ \frac{r_s}{r} - \frac{r_s}{r_0} } 
  - \frac{c}{2}
  \Bigg[
      \sqrt{ \frac{r_s}{r} - \frac{r_s}{r_0} } \times
    \\ &
    \left(
      \frac{s^2}{r^2(1 - r_s/r)}
      +
      \frac{s^2}{r_0^2(1 - r_s/r_0)}
      +
      \frac{s^2}{r r_0(1 - r_s/r)(1 - r_s/r_0)}
    \right) \\
    &+
    \left( \frac{p_q \hbar}{2mcs} \right)^2
    \left( 2 - \frac{r_s}{r} - \frac{r_s}{r_0} \right)
    \sqrt{ \frac{r_s}{r} - \frac{r_s}{r_0} }
    +
    2 \frac{r_s/r}{1 - rs/r} \frac{s}{r} \frac{\dot{s}}{c}
    +
    \frac{1}{\sqrt{ r_s/r - r_s/r_0 }}
    \frac{\dot{s}^2}{c^2}
  \Bigg].
  \end{split}
\end{equation}

In the adiabatic limit we assume that the motion occurs in the \(t-r\) plane and set \(\dot{s}=0\). Then these tangent vector components determine the slope of the massive particle's worldline as
\begin{equation}
  \label{eq:quantumParticleSlope}
  \begin{split}
  \left( \frac{dt}{dr} \right)_{\text{massive particle}} 
  = - \frac{1}{c} &\frac{1}{1 - r_s/r} \sqrt{\frac{r}{r_s}} \sqrt{\frac{1 - r_s/r_0}{1 - r/r_0}} \\
  &\times\Bigg[
    1
     - \frac{1}{2} \left(1 - \frac{r_s}{r}\right)
    \Bigg(
      \Big(
        \frac{1}{(r_0-r_s)^2}
        +
        \frac{1}{(r-r_s)^2}\\
        &+
        \frac{1}{(r_0-r_s)(r-r_s)}
        +
        \frac{(r-r_0)(r+2r_0 - 3 r_s)r_s}{(r-r_s)^3(r_0-r_s)^2}
      \Big) s^2
      +
      \left(
        \frac{p_q\hbar }{2 m c s}
      \right)^2
    \Bigg)
  \Bigg].  
  \end{split}
\end{equation}
The lightcone slope in the \(t-r\) plane is also modified by the perturbed metric functions. 
In the quantum-adiabatic limit where \(ds = 0\), 
\begin{equation}
  \label{eq:quantumLightSlope}
  \left(\frac{dt}{dr}\right)_{\text{lightcone}}
  % =
  % \sqrt{-\frac{g^{\mathcal{Q}}_{rr}}{g^{\mathcal{Q}}_{tt}} - \frac{g^{\mathcal{Q}}_{qq}}{g^{\mathcal{Q}}_{tt}} \left(\frac{\dot{q}}{\dot{r}}\right)^2}
  =
  \pm \frac{1}{c}
    \frac{1}{1-r_s/r}
  \pm
  \frac{r_s s^2}{cr^3 \left(1-r_s/r\right)^3}.
\end{equation}
The ratio of (\ref{eq:quantumParticleSlope}) to (\ref{eq:quantumLightSlope}) yields the quantum generalization of the classical lightcone condition (\ref{eq:classicalLightConeCondition}) as
\begin{equation} \label{causal}
  \begin{split}
    \left \vert \frac{\left( dt/dr \right)_{\text{massive particle}}}{\left( dt/dr \right)_{\text{lightcone}}}
    \right\vert
    = &\sqrt{\frac{r}{r_s}} \sqrt{\frac{1 - r_s/r_0}{1 - r/r_0}} \times \\
    &
      \left(
      1 - \frac{\left(r^2+r (r_0-2 r_s)+(r_0-r_s)^2\right)}{r (r-r_s) (r_0-r_s)^2}\frac{s^2 }{2}
      - \frac{1}{2} \left(\frac{p_q \hbar }{2 m c s}\right)^2 \left(1-\frac{r_s}{r}\right)
      \right).
  \end{split}
\end{equation}

When the massive particle state is localized well away from the horizon, the
ratio of velocities is greater than one, indicating that quantum effects are
consistent with an appropriately generalized notion of causal structure
implied by the extended geometry. However, as the state approaches to within
its correlation distance of the Schwarzschild radius, the particle velocity
can exceed the limit set by the local lightcone. This anomolous behavior was
suggested by the metric signature analysis in Section
\ref{sec:signature-change}.

In this region, a closer analysis including higher
adiabatic orders may shed more light on the dynamics, but a deeper
understanding of space-time structure will likely be required too.  The
signature effects observed when $r$ is close to $r_s$ are caused by quantum
fluctuations of a wave function that, heuristically, is spread out over a
region that crosses the horizon. Even if the radial position is still greater
than the Schwarzschild radius and we may be tempted to consider the
Schwarzschild slicing valid, the support of a wave function includes radii
less than the Schwarzschild radius as well as at this radius itself, where the
coordinate system and the slicing break down. The varying signatures according
to Table~\ref{tab:schwarzschildSignature} are then seen as an implication of a
wave function that simultaneously experiences a spacelike $r$ (outside of the
Schwarzschild radius) and a timelike $r$ (inside). As derived, these effects
become relevant when $r\approx r_s\pm s$, just when the main support of the
wave function reaches the classical position of the horizon. The horizon
structure and, as seen in (\ref{causal}), the causal structure then become
unsharp or, from the perspective of Riemannian geometry, ill-defined.

These observations have an important implication for quantum reference frames
in spacetime. From the formal perspective of reference frames, it would be
allowed to use $t$ as a reference variable for relational evolution if $t$ is
timelike (outside of the horizon) or when $t$ is spacelike (inside), in both
cases a mathematical relationship $r(t)$ is strictly determined by initial
conditions. However, our example demonstrates that quantization of such a
function is not guaranteed to be compatible with causal properties of
spacetime. From the perspective of physical observables, a key problem is that
only the outside $t$ can be related to an observer (an asymptotic one at large
fixed radius) while this observer does not have access to the inside $t$, and
there is no stationary observer in the interior that could measure $t$ there.
It is therefore important to develop a systematic role of spacetime structure
in the general theory of quantum reference frames, at least in cases in which
they are applied to gravitational systems. In this context, it is important to
note perhaps related obstructions to covariance in models of time as a quantum
reference frame with position-dependent momentum terms in the Hamiltonian
constraint \cite{AlgebraicTime,AlgebraicFrozen,TimeFluct}.

A breakdown of covariance does not necessarily constitute an inconsistency in
quantum reference frames. More likely, it would just mean that different
choices of quantum reference frames (such as a stationary clock some distance
from the horizon, compared with a freely falling clock) constitute physically
distinct setups of experiments, and therefore may well predict different
measurement outcomes. The main question is how much of the resulting physics
can still be described geometrically within a meaningful setting of spacetime.

\section{Discussion}
\label{sec:discussion}

In general relativity, time and causality emerge from the geometry of
spacetime itself and are intricately connected. General relativity describes
the universe as a four-dimensional spacetime manifold, with time serving as
one of the coordinate functions that label points. However, the manifold
structure alone does not establish causal relationships. Instead, a
Lorentzian-signature metric tensor \(g_{ab}\) distinguishes between time-like,
space-like, and light-like intervals.  Once the manifold and metric tensor
have been specified, individual notions of time and space are obtained through
metric slicings which preserve its signature.

No matter the slicing, because of the underlying geometric construction, there
must be a mechanism through which observers in different slicing can agree on
the outcomes of experiments. Such fixed outcomes lead to constraints. In the
case of distance measurements, this leads to the spacetime interval having an
invariant  value
\begin{equation}
  \label{eq:5}
  g_{ab} dx^a dx^b = g_{a'b'} dx^{a'} dx^{b'} = \alpha 
\end{equation}
for some fixed constant \(\alpha\). Introducing an arbitrary parametrization promotes this basic constraint to a singular Lagrangian mechanics
\begin{equation}
  \label{eq:7}
  L = \frac{1}{2} g_{ab} \frac{dx^a}{d\sigma} \frac{dx^b}{d\sigma} - \frac{\alpha}{2} = 0.
\end{equation}
We have learned how to translate the singular Lagrangian mechanics into canonical gauge theory governed by a constraint
\begin{equation}
  \label{eq:canonicalGenerator}
  H = \frac{1}{2} g^{ab}p_ap_b + \frac{\alpha}{2} = 0
\end{equation}
In the classical theory, governing dynamics through a constraint is a feature which allows alternative notions of dynamics, generated through alternative slicings, to be equated.

When the metric \(g^{ab}\) is flat, the Hamiltonian (\ref{eq:canonicalGenerator}) may be quickly promoted to a quantum operator
\begin{equation}
  \label{eq:8}
  \hat{H} = - \hat{p}_t^2 + \hat{p}_x^2.
\end{equation}
In this case, one can realize the selection of a time coordinate as a factorization of the Hilbert space into clock and ``system'' spaces, \(\mathcal{H} \cong \mathcal{H}_C \otimes \mathcal{H}_S\). This interprets the canonically-quantized Hamiltonian into the constraint operator
\begin{equation}
  \label{eq:constraintOperator}
  \hat{C} = \hat{H}_C \otimes \pmb{1} + \pmb{1} \otimes \hat{H}_S.
\end{equation}
Physical states satisfy the constraint \(\hat{C} \vert \Psi \rangle \rangle =
0\). Covariant time observables \([ \hat{T}, \hat{H}_c] \propto 1\) with
eigenvectors \(\vert t \rangle\) (or, more generally, POVMs) decompose states as
\begin{equation}
  \label{eq:12}
  \vert \Psi \rangle \rangle = \int dt \vert t \rangle_C \otimes \vert \psi(t) \rangle_S
\end{equation}
where \(\vert \psi(t) \rangle_S = \langle t \vert \Psi \rangle \rangle\). The constraint operator leads to the Schr{\"o}dinger equation for the system state evolving by the system Hamiltonian
\begin{equation}
  \label{eq:13}
  i \hbar \frac{d}{dt} \vert \psi(t) \rangle_S
  = H_S \vert \psi(t) \rangle_S.
\end{equation}
However, the decomposition (\ref{eq:constraintOperator}) is generally not
possible for a metric which depends on the position coordinates. In such a
case, factorization leads to a constraint operator with an additional
clock-system interactions term, \(H_{\text{int}}\). Following the
Page-Wootters mechanism now results in a Schr{\"o}dinger equation in which the
effective Hamiltonian is not Hermitian and the dynamics not unitary \cite{ClockSystemInt}:
\begin{equation}
  \label{eq:13}
  i \hbar \frac{d}{dt} \vert \psi(t) \rangle_S
  = H_S \vert \psi(t) \rangle_S + \int dt' \langle t \vert H_{\text{int}}\vert t'\rangle\vert \psi(t')\rangle_S.
\end{equation}

We presented our phase space quantization as an alternative to the Page-Wootters mechanics for quantization of relativistic systems governed by a constraint.
Canonical effective methods work directly with the observable quantum
statistics \(\langle\hat{x}\rangle,\langle\hat{p}\rangle,\) and \(\Delta(x^ap^a)\). These statistics may be measured and predicted independently of a specific choice of wave function state. 
These methods allowed us to find a description of the quantum dynamics which
formally retains the same geometrical structure of the classical metric
theory. They also led us to highlight important challenges to spacetime
applications of quantum reference frames in the context of covariance. While
our formalism is able to include clock-system interactions implied by curved
spacetime, it also demonstrates a potential breakdown of the classical
spacetime picture and causal concepts such as horizons.

Towards applications, this description gave in Section
\ref{sec:quantum-proper-time} an effective procedure for constructing an
operational notion of quantum proper time in which time retains its geometric
interpretation. The intuitive picture is that of a wavepacket traversing a
metric manifold. However, a quantum wavepacket obeys different equations than
either a classical point particle or even a classical extended object. To
account for this, the metric manifold will not be the original classical
spacetime; rather, it undergoes a deformation contingent upon quantum state
parameters. An open and probably challenging question is whether the
higher-dimensional extension of the metric to quantum degrees of freedom can
be part of a complete covariant formulation. Nevertheless, we demonstrated the
usefulness of this description by our examples, which generalize previous
results from \cite{smith2020} to general semiclassical states in curved
spacetime. They may also be used in other contexts, such as estimating quantum
corrections to time dilation that may be testable by precise lifetime
measurements of unstable particles, such as muons \cite{MuonDilation}.

\section*{Acknowledgements}

\noindent This work was supported in part by NSF grant PHY-2206591.

\appendix

\section{Detailed methods}
\label{a:detailed}

It is possible to work directly from the action in Eq.~(\ref{eq:action}) (see
for instance \cite{gavrilov2000b}), however we prefer to eliminate the square
root at the classical level and work from the so-called einbein action. Following \cite{gavrilov2000a} and \cite{gavrilov2000b}, we include a coupling to a background electromagnetic field primarily to aid later in identifying the physical meaning of the signs of certain terms. That is, we model the particle in background gravitational and electromagnetic fields with the action
\begin{equation}
  \label{eq:SMaster}
  S[x,e] = \int_0^1 d\sigma
  \left(
    - \frac{g_{ab}\dot{x}^a \dot{x}^b}{2e}
    + \frac{e(mc)^2}{2}
    + q\dot{x}^a A_a
  \right),
\end{equation}
where \(A_a\) is the electromagnetic potential, \(q\) is the electric charge
of the particle, and we use the shorthand:
\begin{equation}
  \label{eq:shorthand}
  \dot{x}^a := \frac{dx^a}{d\sigma}.
\end{equation}

The einbein field \(e\) appears in Eq.~(\ref{eq:SMaster}) without derivatives and therefore implies a constraint.
The Lagrangian equations of motion for \(e\) are
\begin{equation}
  \label{eq:constraint}
  \dot{x}^2 + (emc)^2 = 0.
\end{equation}
Solving for the einbein gives
\begin{equation}
  \label{eq:eConstraint}
  e^2 = \frac{-g_{ab}\dot{x}^a\dot{x}^b}{m^2c^2}.
\end{equation}
The Lagrangian equation (\ref{eq:eConstraint}) determines only the magnitude
of \(e\). Its sign \(e\) may be introduced as an independent discrete gauge variable
\begin{equation}
  \label{eq:zeta}
  \zeta := \mathrm{sign}{(e)}.
\end{equation}

Solving for \(e\) from the constraint equation (\ref{eq:eConstraint}) and using the sign parameter \(\zeta\), we can integrate out the einbein field from the action, yielding the reduced action
\begin{equation}
  \label{eq:reducedAction}
  S\left[x,e= \frac{\zeta \sqrt{-\dot{x}^2}}{mc}\right]
  =  \int_0^1 d\sigma
  \left(
    \zeta mc \sqrt{-\dot{x}^2} + q \dot{x}^a A_a
  \right).
\end{equation}

The einbein action, Eq.~(\ref{eq:SMaster}), describes two classes of physical
trajectories, corresponding to each sign of \(\zeta\). Rescaling the reduced
action, Eq.~(\ref{eq:reducedAction}), to move the sign onto the
electromagnetic term suggests that the two signs of the einbein may be
associated with opposite signs of the charge \(q\). In this manner, the
classical theory allows for a simultaneous description of both particles and
antiparticles, a feature explored thoroughly in
Refs.~\cite{gitman1990,gavrilov2000a,gavrilov2000b}.

While antiparticles are indispensable for a consistent quantum theory, our present focus will be to demonstrate the geometric features of the quantum theory. To simplify the analysis, we fix this additional gauge using the condition \(e>0\) or \(\zeta = 1\). Physically, this restricts the theory to the particle region of the one-particle Hilbert space, which is acceptable in the one-particle limit when the sign of the charge \(\zeta\) is conserved. Moving forward, we will also consider neutral particles (or free gravitationally-bound particles) and neglect the electromagnetic coupling.

The Euler-Lagrange equations for \(x\) are
\begin{equation} \label{eq:geodesic}
  \ddot{x}^a + \Gamma_{bc}^a \dot{x}^b \dot{x}^c = \dot{x}^a \frac{\dot{e}}{e}
\end{equation}
where \(\Gamma_{bc}^a\) are the usual Christoffel symbols for the metric \(g_{ab}\).
This expression indicates that a change of parametrization from \(\sigma\) to an affine parameter, \(\tau(\sigma)\), may be accomplished when the einbein field \(e(\sigma)\) solves the first-order inhomogeneous differential equation
\begin{equation}
  \label{eq:affineRelation}
  \frac{de/d\sigma}{e(\sigma)} = - 
  \left(
    \frac{d^2\sigma}{d\tau^2}
  \right)
  \left(
    \frac{d\sigma}{d\tau}
  \right)^{-2}.
\end{equation}
In particular, Eq.~(\ref{eq:affineRelation}) permits a choice where \(e\)
remains constant along the worldline, \(de/d\sigma = 0\) so long as \(\sigma\)
is proportional to \(\tau\), in which case the parameter is affine.

For a massive particle, we can choose the gauge \(e=1/m\), which yields the curved-space particle Lagrange function
\begin{equation}
  L(x^a,\dot{x}^a)
  = -\frac{1}{2} m  g_{ab}\dot{x}^a\dot{x}^b +
  \frac{mc^2}{2} \label{eq:LProperTime}
\end{equation}
 with constraint
\begin{equation}
  \sqrt{-\dot{x}^2} = c. \label{eq:ProperTimeNormalization}
\end{equation}
This gauge choice fixes the parametrization, meaning \(\sigma\) is no longer arbitrary. The derivatives in Eqs.~(\ref{eq:LProperTime}) and (\ref{eq:ProperTimeNormalization}) are taken with respect to proper time.

\subsection{Hamiltonian formulation for geodesics}
\label{sec:hamilt-form}

The einbein action, Eq.~(\ref{eq:SMaster}), (setting \(q=0)\) has the Lagrangian
\begin{equation}
  \label{eq:LMaster}
  L(x^a, e, \dot{x}^a, \dot{e}) =
  - \frac{g_{ab}\dot{x}^a \dot{x}^b}{2e} + \frac{e(mc)^2}{2}.
\end{equation}
Denote generalized Lagrangian coordinates by \(q^i\) for \(i = 0,1,2,3,4\) where \(q^i = x^i\) when \(i = 0,1,2,3\) and \(q^4 = e\). The Hessian matrix of (\ref{eq:LMaster}) is singular,
\begin{equation}
  \label{eq:hessianMatrix}
  \frac{\partial^2 L}{\partial \dot{q}^i\partial\dot{q}^j}
  =
  \begin{pmatrix}
    -\dfrac{g_{ab}}{e} & 0 \\
    0 & 0
  \end{pmatrix}.
\end{equation}
The coordinates are numbered so that the minor of maximum rank is in the upper
left corner. The Hessian matrix in Eq.~(\ref{eq:hessianMatrix}) is already in
block form so the canonical analysis is straightforward. In the language of
\cite{gitman1990}, the velocities of the upper left corner will be
``expressible velocities.'' The remaining velocity is ``inexpressible.'' 
The momenta conjugate to the \(x^a\) are
\begin{equation} \label{eq:xMomentumDef}
  p_a := \frac{\partial L}{\partial \dot{x}^a}
  = -\frac{1}{e} g_{ab} \dot{x}^b
\end{equation}
and the velocities are expressed as
\begin{equation}
  \label{eq:expressibleVelocity}
  \dot{x}^a = -e g^{ab}p_b.
\end{equation}

The nullspace of the Hessian matrix has rank 1 indicating that there is one primary constraint. The primary constraint related to the inexpressible velocity is
\begin{equation} \label{eq:primaryConstraint}
  \Phi^{(1)}(x^a,e,p_a,p_e) = p_e := \frac{\partial L}{\partial \dot{e}} = 0 .
\end{equation}

We denote the momenta conjugate to the \(q^i\) by \(p_i\) with \(i = 0,1,2,3,4\) and sums over the index \(i\) will be assumed to be over its entire range. We construct the canonical Hamiltonian from the Legendre transform of the Lagrangian:
\begin{equation} \label{eq:legendreTransform}
  H(q^i, p_i)
  =
  \dot{q}^i p_i - L.
\end{equation}
Substituting Eq.~(\ref{eq:expressibleVelocity}) for the expressible velocities yields the canonical Hamiltonian
\begin{equation}
  H(x^a,p_a, e, p_e) =
  - \frac{e}{2}
    \left(
    g^{ab}p_a p_b + (mc)^2
    \right). 
\end{equation}
The primary Hamiltonian is obtained by adding the primary constraint with a Lagrange multiplier. In this case, following from the equations of motion, we identify the Lagrange multiplier of the primary constraint \(\Phi^{(1)} = p_e\) as the inexpressible velocity \(\dot{e}\) obtaining,
\begin{equation}
  \label{eq:classicalGeneralHamiltonFunction}
  H(x^a,p_a, e, p_e)
  = p_e \dot{e} - \frac{e}{2}
    \left(
    g^{ab}p_a p_b + (mc)^2
    \right).
\end{equation}

The consistency condition enforcing conservation of the primary constraint introduces a secondary constraint
\begin{equation}
  \label{eq:massshellConstraint}
  0 = \dot{p}_e
  = [p_e, H] 
  = \frac{1}{2}
    \left(
    g^{ab}p_a p_b + (mc)^2
    \right) =: \Phi^{(2)}(x,p)
\end{equation}

This is the mass-shell constraint. In the einbein action, Eq.~(\ref{eq:SMaster}), this constraint is secondary because it does not follow immediately from the momentum definitions; instead it relies on the Lagrangian equation of motion, Eq.~(\ref{eq:eConstraint}).
However, the same constraint appears as a primary constraint of the square-root Lagrangian, Eq.~(\ref{eq:action}). Because of this and the fact that both constraints \(\Phi^{(1)}\) and \(\Phi^{(2)}\) are first class, we assume both constraints generate gauge transformations. This interpretation is consistent with the Dirac conjecture discussed in \cite{brown2022}.

The consistency condition \(\dot{\Phi}^{(2)} = 0\) gives no further constraints nor conditions on the Lagrange multipliers.
The first-order Hamiltonian action is then
\begin{equation}
  \label{eq:3}
  S[ x^a, e, p_a, p_e, \lambda^1, \lambda^2 ]
  =
  \int d\tau
  \left[
    p_a \dot{x}^a + p_e \dot{e} - H - \lambda^1 \Phi^{(1)} - \lambda^2 \Phi^{(2)}
  \right]
\end{equation}
where \(H\) is a first-class Hamilton function and \(\lambda^1\) and \(\lambda^2\) are Lagrange multipliers. 

The equations of motion of the Hamiltonian action obtained from the requirement \(\delta S = 0\) for arbitrary variations of its arguments are
\begin{align}
  \frac{\delta S}{\delta x^a} = 0 \quad &\iff \quad
  \dot{p}_a = \frac{e - \lambda^2}{2} \frac{\partial g^{bc}}{\partial x^a} p_b p_c \\
  \frac{\delta S}{\delta p_a} = 0 \quad &\iff \quad
  \dot{x}^a = - (e - \lambda^2) g^{ab}p_b \\
  \frac{\delta S}{\delta e} = 0 \quad &\iff \quad
  \dot{p}_e = \frac{1}{2} (g^{ab} p_ap_b + m^2c^2) \\
  \frac{\delta S}{\delta p_e} = 0 \quad &\iff \quad
  \dot{e} = \lambda^1 \label{eq:eHamiltonian} \\
  \frac{\delta S}{\delta \lambda^1} = 0 \quad &\iff \quad
  p_e = 0 \\
  \frac{\delta S}{\delta \lambda^2} = 0 \quad &\iff \quad
  \frac{1}{2} (g^{ab} p_ap_b + m^2c^2) = 0
\end{align}
The constrained Hamiltonian dynamics represented by these equations are equivalent to the Lagrangian equations (\ref{eq:eConstraint}) and (\ref{eq:geodesic}).

The solutions depend on two arbitrary functions, \(\lambda^1\) and \(\lambda^2\). The solution to Eq.~(\ref{eq:eHamiltonian}) is immediate,
\begin{equation}
  \label{eq:4}
  e(\sigma) = \int \lambda^1(\sigma') d\sigma'.
\end{equation}
This result implies that the einbein field \(e\) is arbitrary.

The phase space is ten dimensional, \(\Omega = (\mathbb{R}^{10}, \omega)\), with the canonical symplectic form
  \begin{equation}
    \label{eq:48}
    \omega = de \wedge dp_e + dx^a \wedge dp_a.
  \end{equation}
The phase-space function  \(\Phi^{(1)} = p_e\) is the Hamiltonian generator of gauge transformations. 
  The Hamiltonian vector field of the gauge constraint, \(G\), is obtained by treating \(p_e\) as the Hamilton function,
  \begin{equation}
    \label{eq:49}
    i_G \omega = dp_e
  \end{equation}
  and we find
  \begin{equation}
    \label{eq:50}
    G = \frac{\partial}{\partial e}.
  \end{equation}
  The integral curves of \(G\) are obtained from the flow with parameter \(\sigma\), \(\Phi_G^\sigma\), from the differential equations
  \begin{equation}
    \label{eq:43}
    \frac{d}{d\sigma} \Phi_G^\sigma = G \circ \Phi_G^\sigma.
  \end{equation}
  \begin{align}
    L_G(x^a) = 0 \quad &\implies \quad x^a(\sigma) = x^a(0) \\
    L_G(e) = 1 \quad &\implies  \quad e(\sigma) = e(0) + \sigma \\
    L_G(p_a) = 0 \quad &\implies \quad  p_a(\sigma) = p_a(0) \\
    L_G(p_e) = 0 \quad &\implies \quad  p_e(\sigma) = p_e(0)\,.
  \end{align}
Flowing along \(G\) therefore connects configurations with different values of the einbein field.

In the equations of motion the Lagrange multiplier \(\lambda^2\) appears only in the combination \(e-\lambda^2\). We can exploit the gauge freedom in \(\lambda^1\) to choose \(e\) such that \(\lambda^2 = 0\).
The remaining equations of motion are
\begin{align}
  \frac{1}{e} \frac{d x^a}{d\sigma}
  &=  -  g^{ab} p_b \\
  \frac{1}{e} \frac{d p_a}{d \sigma}
  &=
  \frac{1}{2}
  \left(\partial_a g^{bc}\right) p_b p_c.
\end{align}
The field \(e\) may be absorbed into the derivative, defining a new parameter function. Alternatively we can choose the gauge condition for a massive particle
\begin{equation}
  \label{eq:39}
  \Phi^{(3)} = e - \frac{1}{m} = 0.
\end{equation}

This gauge condition, together with the constraint \(\Phi^{(1)}\), form a second-class constraint algebra. Imposing these constraints allows us to eliminate \(e\) and \(p_e\) from the theory. As a result, we obtain the partially-reduced Hamiltonian
\begin{equation}
  \label{eq:classicalProperTimeHamiltonFunction2}
  H(x^a,p_a)
  = - \frac{1}{2m}
  \left(
    g^{ab}p_a p_b + (mc)^2
  \right).
\end{equation}
We call this the proper-time Hamiltonian. We leave the remaining constraint, \(\Phi^{(2)}\), unfixed and proceed immediately to canonical quantization via Dirac's quantization of a Hamiltonian theory with constraints \cite{dirac1964}.

\subsection{Canonical quantization}
\label{sec:quantization}

In canonical quantization, the quantum system is specified by a unital algebra of observables \(\mathcal{A}\) whose commutation relations are derived from the classical theory. 
In our application, the nonzero commutators for the algebra's generators are
\begin{equation}
  \label{eq:canonicalCommutationRelations}
  [ \hat{x}^a, \hat{p}_b] = i \hbar \delta^a_b\,.
\end{equation}
A representation of these relations on the Hilbert space \(\mathcal{H} = L^2(\mathbb{R}^4)\) with inner-product
\begin{equation}
  \label{eq:42}
  \langle \phi, \psi \rangle
  :=
  \int \overline{\phi(x)} \psi(x) d^4x
\end{equation}
is realized as
\begin{equation}
  \label{eq:54}
  \hat{x}^a \rightarrow x^a \qquad
  \hat{p}_a \rightarrow \frac{\hbar}{i} \frac{\partial}{\partial x^a}.
\end{equation}

Unitary dynamics arise from the time-evolution operator
\begin{equation}
  \label{eq:55}
  U(\tau) = \exp \left( - \frac{i}{\hbar} \hat{H} \tau \right)
\end{equation}
where, adopting the classically-defined proper-time gauge, canonical quantization of the curved-space Hamilton function (\ref{eq:classicalProperTimeHamiltonFunction}) gives for the quantum system
\begin{equation} \label{eq:quantumHamiltonian}
  \hat{H}(\hat{x},\hat{p}) = -\frac{1}{2m} \left( g^{ab}(\hat{x}) \hat{p}_a \hat{p}_b + m^2c^2I \right)
  + \mathrm{h.c.}
\end{equation}
with the identity operator $I$. In general, the choice of a symmetric ordering
is required in order to turn this operator into an unambiguous Hermitian expression.

Because the classical theory was constrained, the physical Hilbert space
should be defined as the Cauchy completion of a suitable set of states annihilated by the Hamiltonian constraint:
\begin{equation}
  \label{eq:23}
  \mathcal{D}_{\text{phys}} = \{\psi \in L^2(\mathbb{R}^4) : \hat{H} \psi = 0 \}.
\end{equation}
If zero is in the discrete spectrum of $\hat{H}$, the solution space is
contained in the original Hilbert space $\mathcal{H}$ and formd the physical Hilbert space
after Cauchy completion. If zero is in the continuous part of the spectrum,
$\mathcal{D}_{\text{phys}}$ is not a subset of $\mathcal{H}$ but can be
interpreted as a set of distributional states on $\mathcal{H}$. This set
should then be equipped with an inner product, such that its Cauchy completion
forms the physical Hilbert space. For details and a specific example related
to the relativistic particle, see \cite{GenRepIn}.

The quantum theory so far outlined has several challenges. For one, the
appropriate ordering of operators in Eq.~(\ref{eq:quantumHamiltonian})
presents an inherent ambiguity, and further choices are usually required in
the definition of a physical inner product.  Secondly, the Hamiltonian
constraint defining \(\mathcal{H}_{\text{phys}}\) results in states which do
not depend on proper time. Lastly, owing to the metric operator in
Eq.~(\ref{eq:quantumHamiltonian}), this Hamiltonian is not quadratic. 

Each of these challenges can be overcome or evaded, at least for semiclassical
states, by the reformulation of the Hilbert space theory  we present below,
based on \cite{EffCons,EffConsRel}.
The ordering ambiguity will be addressed by expanding \(\hat{H}\) in a series whose individual terms are easily symmetrized.
The solvability of the system is addressed by truncating this series. Lastly,
we will see that the reformulation allows the constrained quantum theory to be re-interpreted as a constrained theory of classical Hamiltonian type. Identifying the quantum state degrees of freedom in this language allows us to extend immediately the classical geometric notion of time as arc length, eliminating the need for a novel solution to the time evolution problem.

\subsection{Geometrization}
\label{sec:canon-effect-meth}

So far we have introduced a Hilbert space formulation of the quantum theory based on an operator algebra derived from the classical theory and a choice of representation for this algebra. We now develop our geometric formulation of the quantum theory through a careful consideration of the notion of quantum state.

In modern quantum theory there are several alternative notions of quantum
state. On the one hand, taking the algebra of observables as primary, a
quantum state can be defined as a linear functional
\(\omega\colon \mathcal{A} \rightarrow \mathbb{C}\). Alternative to the
algebraic approach, the state of a quantum system can be defined as a
positive, trace-class and trace-one, linear operator $\rho$ on a Hilbert space
\(\mathcal{H}\) that carries an irreducible representation of the algebra
$\mathcal{A}$. In this case of a density matrix, we have
$\omega(\hat{A})=\mathrm{Tr}(\rho \hat{A})$ for
$\hat{A}\in\mathcal{A}$. The purely algebraic formulation is more general
because it does not require the introduction of a Hilbert space, provided the
algebra is equipped with a star-structure that generalizes adjointness
relations of operators. In either case, pure states play an important role in
the theory because they represent the most precise specification of a quantum
state. One typically denotes a pure state operator \(\rho_\psi\) in terms of a
representative \(\psi \in \mathcal{H}\) as
\(\rho_\psi = \vert \psi \rangle \langle \psi \vert/\langle \psi \vert \psi
\rangle\).

Pure states are insensitive to arbitrary complex (and possibly time dependent) rescaling of the representative. This motivates us to define an equivalence relation \(\sim \) on the space of pure quantum states such that two states are equivalent if and only if they are related by complex rescaling of the representative:
\begin{equation}\label{rescale}
  \rho_{\lambda(\tau) \psi} \sim \rho_\psi \quad:\iff \quad \lambda(\tau) \in \mathbb{C} \text{ at fixed } \tau.
\end{equation}

The quotient of the Hilbert space by this equivalence relation defines the projective Hilbert space \(P(\mathcal{H}) := \mathcal{H}/\sim\). Elements of the projective Hilbert space can be identified one-to-one with the pure states of the quantum system. By removing the scaling redundancy, the projective Hilbert space provides a more precise structure to work with than the Hilbert space itself.

For finite dimensional systems, the projective Hilbert space is isomorphic to finite dimensional complex projective spaces. For example, for the two-dimensional spin--1/2 system, the space of all pure states is the Bloch sphere. However, whenever \(\mathcal{H}\) is infinite-dimensional the projective Hilbert space is also infinite dimensional.

The geometric properties of the projective Hilbert space have been extensively studied \cite{bjelakovic2005, strocchi1966,kibble1979, ashtekar1999} and provide useful language for working with this space. In particular, the projective Hilbert space always has the structure of a K{\"a}hler manifold. The K{\"a}hler structure is derived from the decomposition of the Hilbert space inner product into real and imaginary parts as
\begin{equation}
  \label{eq:40}
  \langle \phi, \psi \rangle = \frac{1}{2\hbar} G(\phi, \psi) + \frac{i}{2\hbar} \Omega(\phi, \psi)
\end{equation}
where \((2\hbar)^{-1} G\) and \((2\hbar)^{-1}\Omega\) represent the real and imaginary parts, respectively. It follows from properties of the sesquilinear inner product that \(G\) is a Riemannian metric while \(\Omega\) is a symplectic form, with the two related by the complex structure
\begin{equation}
  \label{eq:41}
  G(\phi, \psi) = \Omega(\phi, i \psi).
\end{equation}

We refer to the projective Hilbert space \(P(\mathcal{H})\) as the quantum phase space because of its symplectic structure. The symplectic structure can be expressed through a Poisson bracket.
Given arbitrary operators $\hat{A}$ and $\hat{B}$ acting on the Hilbert space, the expectation values $g_{\hat{A}}(\psi)=\langle\psi|\hat{A}|\psi\rangle / \langle \psi \vert \psi \rangle$ and $g_{\hat{B}}(\psi)=\langle\psi|\hat{B}|\psi\rangle /\langle \psi \vert \psi \rangle$ can be interpreted as functions on the projective Hilbert space because they depend on the state $\psi$ in which they are computed and have unique values on each ray of the Hilbert space.
The Poisson bracket of these functions is defined by
\begin{equation}
  \{g_{\hat{A}},g_{\hat{B}}\}
  =
  \frac{1}{i\hbar}
  \frac{ \langle \psi \vert [ \hat{A}, \hat{B}] \vert \psi \rangle }
  {\langle \psi \vert \psi \rangle}
  = \frac{1}{i\hbar}g_{[\hat{A},\hat{B}]}\,.
  \label{eq:poissonBracket}
\end{equation}

It is possible to rewrite this Poisson bracket explicitly as an expression between expectations values of pure states as
\begin{equation}
  \label{eq:1}
  \left\{ \mathrm{Tr}(\rho_\psi \hat{A}),
    \mathrm{Tr}(\rho_\psi \hat{B})
  \right \}
  = \frac{1}{i\hbar} \mathrm{Tr}(\rho_\psi [\hat{A},\hat{B}]).
\end{equation}
However, the relationship with (\ref{eq:poissonBracket}) breaks down  when
\(\rho_\psi\) is replaced by a general mixed state \(\rho\). Nevertheless,
(\ref{eq:1}) fulfills all the requirements for a Poisson bracket on the space
of mixed states, if the Leibniz rule for products of expectation values is
included in the definition. The inclusion of mixed states therefore requires a
generalization of the constructions given in \cite{bjelakovic2005,
  strocchi1966,kibble1979, ashtekar1999}, which can still be formulated
by means of a Poisson bracket \cite{bojowald2006} on the space of states,  but
in general not with a K\"ahler structure.

\subsubsection{Coordinates}

Points \(\rho_\psi \in P(\mathcal{H})\) of the manifold are pure quantum states. One possible choice of coordinates on the manifold \(P(\mathcal{H})\) are those given by the statistics corresponding to the state.
These coordinates are constructed as functions on \(P(\mathcal{H})\) beginning with the basic expectation values
\begin{align}
  x^a &\equiv \langle \psi \vert \hat{x}^a \vert \psi \rangle / \langle \psi \vert \psi \rangle \\
  p_a &\equiv \langle \psi \vert \hat{p}_a \vert \psi \rangle / \langle \psi \vert \psi \rangle
\end{align}
where \(a = 0,1,2,3\) label the spacetime degrees of freedom.
We select the remaining coordinates to be the centralized and symmetrized higher statistical moments---including correlations and fluctuations---defined as
\begin{equation}
  \Delta \big ((x^0)^{\alpha_0}\cdots (x^3)^{\alpha_3}p_0^{\beta_0}\cdots p_3^{\beta_3} \big)
  :=\frac{
  \big\langle \psi \vert
  (\hat{x}^0-x^0)^{(\alpha_0}\cdots(\hat{x}^3-x^3)^{(\alpha_3}(\hat{p}_0-p_0)^{\beta_0)}\cdots(\hat{p}_3-p_3)^{\beta_3)}
  \vert \psi
  \big \rangle }{ \langle \psi \vert \psi \rangle}
\end{equation}
where parentheses in the superscript indices indicate symmetrization. Also
this set of coordinates can be generalized to mixed states if the trace is
used to compute expectation values.

We can rewrite the higher moments more succinctly by introducing the multi-indices \(\alpha = (\alpha_0,\alpha_1,\alpha_2, \alpha_3)\) and \(\beta = (\beta_0, \beta_1,\beta_2,\beta_3)\). These are defined to expand over the spacetime degrees of freedom as
\begin{align}
  x^\alpha = (x^0)^{\alpha_0} (x^1)^{\alpha_1} (x^2)^{\alpha_2} (x^3)^{\alpha_3} \\
  p^\beta = (p_0)^{\beta_0} (p_1)^{\beta_1} (p_2)^{\beta_2} (p_3)^{\beta_3}.
\end{align}
Then we can write a general higher quantum moment as
\begin{equation}
  \Delta \big ((x^0)^{\alpha_0}\cdots (x^3)^{\alpha_3}p_0^{\beta_0}\cdots p_3^{\beta_3} \big)
  \equiv \Delta \big (x^{\alpha}p^{\beta}\big).
\end{equation}
Because these functions form a coordinate chart over quantum phase space, we
can represent a pure or mixed state uniquely by its infinite set of quantum moments:
\begin{equation}
  \label{eq:2}
  \rho_\psi \leftrightarrow \left( x^a, p_a, \Delta(x^\alpha p^\beta) \right)
  \text{ for all multi-indices }\alpha, \beta.
\end{equation}

Equivalence in the forward direction follows from the definition of the
quantum moments. The reverse equivalence---the reconstruction of a unique
quantum distribution from a collection of pre-specified statistics, known as
the moment problem---is more challenging to prove rigorously. For discussion
of the state reconstruction problem related to the methods used here see \cite{bojowald2006} and
\cite{balsells2023}.

We define further arithmetic rules for the multi-indices. The multi-indices should distribute over the factorial operation as
\begin{equation}
  \label{eq:45}
  \alpha! := \alpha_0! \alpha_1! \alpha_2 ! \alpha_3!\,.
\end{equation}
The order of a multi-index is computed by summing the entries and denoted
\begin{equation}
  \label{eq:47}
  \vert \alpha \vert =
  \alpha_0 + \alpha_1 + \alpha_2 + \alpha_3\,.
\end{equation}
The order of a moment is defined as the value of the sum \(\vert \alpha + \beta \vert = \sum_i (\alpha_i + \beta_i)\). 
A state specified by these statistical observables is said to be semiclassical if its moments satisfy the hierarchy condition
\begin{equation}
  \label{eq:hierarchy2}
  \Delta(x^\alpha p^\beta) = O(\hbar^{\vert \alpha + \beta \vert /2}).
\end{equation}
Such conditions are satisfied for Gaussian states, but also by more general states because the specific coefficients of $\hbar^{\sum_i(\alpha_i+\beta_i)/2}$ are not determined by the condition \eqref{eq:hierarchy2}. The hierarchy condition captures the semiclassical limit in the sense that states obeying Eq.~\eqref{eq:hierarchy2} will be concentrated around their mean values with fluctuations that decrease in magnitude at higher orders.

\subsubsection{Dynamics}

The Poisson structure allows us to associate to each observable a vector field generating a Hamiltonian flow on the (quantum) phase space. We are especially interested in the dynamics obtained from the quantum Hamilton function defined as
\begin{equation}
  \label{eq:51}
  H_{\mathcal{Q}}(\rho_\psi) := {\rm Tr}(\rho  \hat{H})
\end{equation}
or as the straightforward extension to mixed states.

By definition, the quantum Hamilton function is the function on quantum phase space obtained by evaluating the expectation value of the Hamiltonian operator in a generic state. In the relativistic case, this function is constrained to vanish. However, when states are parameterized by their basic expectation values, $x = \langle\hat{x}\rangle$ and $p = \langle\hat{p}\rangle$, along with central moments \(\Delta(x^\alpha p^\beta)\), $H_{\mathcal{Q}}$ becomes a function of these variables and it is useful to develop a general expression for \(H_{\mathcal{Q}}\) showing its dependence on these phase space variables. Such an expression can be obtained through a series expansion centered around the basic expectation values:
\begin{align}
  H_\mathcal{Q}\big (x^a,p_a,\Delta(x^\alpha p^\beta) \big )
  &\equiv {\rm Tr}\big(\rho
    \hat{H} \big (x^a + (\hat{x}^a - x^a),p_a + (\hat{p}_a - p_a)\big )\big)
    \nonumber \\[1em] 
  &= \sum_{\alpha=0}^\infty\sum_{\beta=0}^\infty
    \frac{1}{\alpha!\beta !} \frac{\partial^{\vert \alpha+\beta \vert }H(x,p)}{\partial^{\alpha}x\partial^{\beta}p} \; \Delta(x^\alpha p^\beta) \label{eq:Hq}
\end{align}
where, allowing for some abuse of notation, the sums are over all multi-indices beginning from the zero multi-index \(0= (0,0,0,0)\). In differential geometric terms, Eq.~(\ref{eq:Hq}) is the coordinate representation of the quantum Hamilton function.

Terms in the sum in Eq.~(\ref{eq:Hq}) are distinguished by the order of the quantum moments they contain. The zeroth order term reproduces exactly the classical Hamilton function. One can check that, owing to their definition as central moments, the first order term in the sum vanishes identically. 
With the quantum corrections starting at second order in \(\sqrt{\hbar}\) (or an appropriate dimensionless quantity), additional terms in the quantum Hamilton function, Eq.~(\ref{eq:Hq}), modify the classical dynamics, leading to a more complete description of the system's behavior. The full dynamics of the system can be understood by analyzing how the sum in Eq.~(\ref{eq:Hq}) generates evolution in the quantum phase space, incorporating both classical trajectories and quantum effects.

\subsubsection{Gauge transformations and constraints}

If \(A(x,p, \Delta(x^\alpha p^\beta))\) is a dynamical variable (explicitly
\(\tau\)-independent), the quantum Hamilton function generates a flow on the phase space according to the phase space structure via
\begin{equation} \label{eq:dynamics}
  \frac{{\rm d}}{{\rm d}\tau} A = \{A,H_{\mathcal{Q}}\}.
\end{equation}
Since the quantum Hamilton function represents a constraint, the evolution it
generates is interpreted as a gauge transformation on the phase space. This
reformulation recasts the constrained quantum theory as a classical-type
Hamiltonian gauge theory. The gauge flow is generated by the phase-space
function $H_{\rm machcal{Q}}=\langle\hat{H}\rangle$, which is constrained to
vanish on physical states. In addition, any expression of the form $\langle
\hat{O}\hat{H}\rangle$ with a generic operator $\hat{O}$ polynomial in
$\hat{x}$ and $\hat{p}$ has a vanishing expectation value in physical
states. As phase-space functions on non-physical states,
$\langle\hat{H}\rangle$ and $\langle
\hat{O}\hat{H}\rangle$ are in general independent functions in the sense that
$\langle\hat{H}\rangle=0$ does not imply $\langle
\hat{O}\hat{H}\rangle=0$. Moments of physical states are therefore obtained
only if we impose a set of constraints, given by $H_{\mathcal{Q}}=0$ as well
as higher-order constraints $\langle
\hat{O}\hat{H}\rangle=0$ for suitable $\hat{O}$, such as all polynomials up to
a certain degree (if we use a truncation) in the basic
operators. Heuristically, the first constraint eliminates unphysical
expectation values among the original phase-space degrees of freedom, while
higher-order constraints eliminate their moments. In addition, all these
constraints generate gauge transformations.

Equation (\ref{eq:dynamics}) is equivalent to the generalized Ehrenfest
theorem. When applying this equation to an observable, the system of coupled
differential equations resulting from repeated use of Eq.~(\ref{eq:dynamics})
will generally fail to close. However, in the semi-classical limit we obtain a
closed dynamical system by truncating terms above a fixed order. Higher-order
constraints should then be truncated at corresponding orders. A detailed
analysis of such gauge systems, given in \cite{EffCons,EffConsRel}, shows that
the resulting constraints are mutually consistent in the sense that all gauge
transformations respect the joint constraint surface on which all the
constraints up to a given order are zero. Moreover, at least for Hamiltonians
that do not depend explicitly on time, a consistent gauge fixing can be chosen
that sets moments involving a time variable, given here by coordinate time $t$, equal to
zero. The resulting reduced phase space on which constraints are solved with
this gauge fixing is then equivalent to a semiclassical quantization of the
classically reduced system.

Using these general results, we can, as a matter of modelling, choose to
quantize only some of the degrees of freedom left after fixing the gauge. In our applications we will
consider radial spacetime trajectories which have a classical phase space with
two canonical pairs of degrees of freedom coordinatized by \((t,r,p_t,p_r)\).
For two degrees of freedom there are 4 basic expectation values and 10
second-order moments.  The 10 second order moments are
\begin{equation}
  \left\{\Delta \left(t^2\right),\Delta \left(r^2\right),\Delta \left(p_t^2\right),\Delta
   \left(p_r^2\right),\Delta (r t),\Delta (p_t t),\Delta (p_r t),\Delta
   (p_t r),\Delta (p_r r),\Delta (p_r p_t)\right\}.
\end{equation}

The system defined by these degrees of freedom has many interesting features,
including relationships between position, energy, and time dispersions, as
well as their correlations. We consider in greater depth the full richness of
this system in a follow-up paper. In this paper, we treat the moments
\(\Delta(t^2), \Delta(tp_t), \Delta(p_t^2), \Delta(tr), \Delta(tp_r),
\Delta(rp_t)\), and \(\Delta(p_tp_r)\) as negligible. This assumption is
equivalent to perturbing only the radial degree of freedom, and it is
justified by a detailed combination of gauge-fixing conditions and solutions
of higher-order constraints. Reducing the phase space dimension allows us to treat the quantum feature of spatial delocation independently and more clearly.

\subsubsection{Symmetries}

An essential feature of the classical theory is its covariance under spacetime
transformations. However, the decision to fix the gauge by restricting moments involving the
time coordinate is a frame-dependent condition, just as the choice of a
classical time coordinate is a choice of frame. In the context of our quantum
framework, we now explore how the dynamics generated by the quantum Hamilton
function remain covariant under spacetime transformations, extending the
classical notion of symmetry to the quantum regime.

In the classical theory, we describe a local change of coordinates by the Lorentz transformation
\begin{equation}
  \label{eq:44}
  x^{a'} = \Lambda^{a'}_{\, a} x^a.
\end{equation}
This transformation gives in the quantum theory the corresponding unitary transformation of the operators
\begin{equation}
  \label{eq:44}
  U(\Lambda)^{-1} \hat{x}^a U(\Lambda) = \Lambda^{a'}_{\, a} \hat{x}^a =: \hat{x}^{a'}.
\end{equation}
By appropriate insertions of the identity, the quantum statistics transform as tensor representations of the Lorentz group. For example, the second-order correlations have the transformation laws
\begin{align}
  \label{eq:secondOrderTransformxx}
  \Delta 
  \left(    x^{a'} x^{b'}  \right)
  &=
  \Lambda^{a'}_{\enspace a}
  \Lambda^{b'}_{\enspace b'}
  \Delta 
  \left(    x^{a} x^{b}  \right)
  \\
  \label{eq:secondOrderTransformxp}
    \Delta 
  \left(    x^{a'} p_{b'}  \right)
  &=
  \Lambda^{a'}_{\enspace a}
  \Lambda^{\enspace b}_{b'}
  \Delta 
  \left(    x^{a} p_{b}  \right)
  \\
  \label{eq:secondOrderTransformpp}
    \Delta 
  \left(    p_{a'} p_{b'}  \right)
  &=
  \Lambda^{\enspace a}_{a'}
  \Lambda^{\enspace b}_{b'}
  \Delta 
  \left(    p_{a} p_{b}  \right) .
\end{align}
With these transformation laws, the moment-based quantum theory is covariant
under local changes of coordinates.  The transformation laws,
Eqs.~(\ref{eq:secondOrderTransformxx})--(\ref{eq:secondOrderTransformpp}), mix
time and space statistics. Neglecting moments involving the time coordinate is
a frame-dependent assumption on the quantum state.

%\bibliographystyle{apsrev}
%\bibliography{references}

\begin{thebibliography}{47}
\expandafter\ifx\csname natexlab\endcsname\relax\def\natexlab#1{#1}\fi
\expandafter\ifx\csname bibnamefont\endcsname\relax
  \def\bibnamefont#1{#1}\fi
\expandafter\ifx\csname bibfnamefont\endcsname\relax
  \def\bibfnamefont#1{#1}\fi
\expandafter\ifx\csname citenamefont\endcsname\relax
  \def\citenamefont#1{#1}\fi
\expandafter\ifx\csname url\endcsname\relax
  \def\url#1{\texttt{#1}}\fi
\expandafter\ifx\csname urlprefix\endcsname\relax\def\urlprefix{URL }\fi
\providecommand{\bibinfo}[2]{#2}
\providecommand{\eprint}[2][]{\url{#2}}

\bibitem[{\citenamefont{Rovelli}(2008)}]{rovelli2008}
\bibinfo{author}{\bibfnamefont{C.}~\bibnamefont{Rovelli}},
  \bibinfo{journal}{Living Reviews in Relativity}
  \textbf{\bibinfo{volume}{11}}, \bibinfo{pages}{5} (\bibinfo{year}{2008}),
  ISSN \bibinfo{issn}{1433-8351},
  \urlprefix\url{https://doi.org/10.12942/lrr-2008-5}.

\bibitem[{\citenamefont{Ali and Engli\v{s}}(2005)}]{ali2005}
\bibinfo{author}{\bibfnamefont{S.~T.} \bibnamefont{Ali}} \bibnamefont{and}
  \bibinfo{author}{\bibfnamefont{M.}~\bibnamefont{Engli\v{s}}},
  \bibinfo{journal}{Reviews in Mathematical Physics}
  \textbf{\bibinfo{volume}{17}}, \bibinfo{pages}{391} (\bibinfo{year}{2005}),
  \eprint{https://doi.org/10.1142/S0129055X05002376},
  \urlprefix\url{https://doi.org/10.1142/S0129055X05002376}.

\bibitem[{\citenamefont{Douglas}(2019)}]{douglas2019}
\bibinfo{author}{\bibfnamefont{M.~R.} \bibnamefont{Douglas}},
  \bibinfo{journal}{Universe} \textbf{\bibinfo{volume}{5}}
  (\bibinfo{year}{2019}), ISSN \bibinfo{issn}{2218-1997},
  \urlprefix\url{https://www.mdpi.com/2218-1997/5/7/176}.

\bibitem[{\citenamefont{Hossenfelder}(2013)}]{hossenfelder2013}
\bibinfo{author}{\bibfnamefont{S.}~\bibnamefont{Hossenfelder}},
  \bibinfo{journal}{Living Reviews in Relativity}
  \textbf{\bibinfo{volume}{16}}, \bibinfo{pages}{2} (\bibinfo{year}{2013}),
  ISSN \bibinfo{issn}{1433-8351},
  \urlprefix\url{https://doi.org/10.12942/lrr-2013-2}.

\bibitem[{\citenamefont{Zych and Brukner}(2018)}]{zych2018}
\bibinfo{author}{\bibfnamefont{M.}~\bibnamefont{Zych}} \bibnamefont{and}
  \bibinfo{author}{\bibfnamefont{{\v{C}}.}~\bibnamefont{Brukner}},
  \bibinfo{journal}{Nature Physics} \textbf{\bibinfo{volume}{14}},
  \bibinfo{pages}{1027} (\bibinfo{year}{2018}), ISSN \bibinfo{issn}{1745-2481},
  \urlprefix\url{https://doi.org/10.1038/s41567-018-0197-6}.

\bibitem[{\citenamefont{Polchinski}(2017)}]{polchinski2017}
\bibinfo{author}{\bibfnamefont{J.}~\bibnamefont{Polchinski}},
  \emph{\bibinfo{title}{The Black Hole Information Problem}}
  (\bibinfo{publisher}{WORLD SCIENTIFIC}, \bibinfo{year}{2017}), pp.
  \bibinfo{pages}{353--397}, New Frontiers in Fields and Strings, ISBN
  \bibinfo{isbn}{978-981-314-943-4}.

\bibitem[{\citenamefont{Martin}(2012)}]{martin2012}
\bibinfo{author}{\bibfnamefont{J.}~\bibnamefont{Martin}},
  \bibinfo{journal}{Comptes Rendus. Physique} \textbf{\bibinfo{volume}{13}},
  \bibinfo{pages}{566} (\bibinfo{year}{2012}),
  \urlprefix\url{https://comptes-rendus.academie-sciences.fr/physique/articles/10.1016/j.crhy.2012.04.008/}.

\bibitem[{\citenamefont{Kucha\v{r}}(1992)}]{KucharTime}
\bibinfo{author}{\bibfnamefont{K.~V.} \bibnamefont{Kucha\v{r}}}, in
  \emph{\bibinfo{booktitle}{Proceedings of the 4th Canadian Conference on
  General Relativity and Relativistic Astrophysics}}, edited by
  \bibinfo{editor}{\bibfnamefont{G.}~\bibnamefont{Kunstatter}},
  \bibinfo{editor}{\bibfnamefont{D.~E.} \bibnamefont{Vincent}},
  \bibnamefont{and} \bibinfo{editor}{\bibfnamefont{J.~G.}
  \bibnamefont{Williams}} (\bibinfo{publisher}{World Scientific},
  \bibinfo{address}{Singapore}, \bibinfo{year}{1992}).

\bibitem[{\citenamefont{Anderson}(2012)}]{anderson2012}
\bibinfo{author}{\bibfnamefont{E.}~\bibnamefont{Anderson}},
  \bibinfo{journal}{Annalen der Physik} \textbf{\bibinfo{volume}{524}},
  \bibinfo{pages}{757} (\bibinfo{year}{2012}),
  \eprint{https://onlinelibrary.wiley.com/doi/pdf/10.1002/andp.201200147},
  \urlprefix\url{https://onlinelibrary.wiley.com/doi/abs/10.1002/andp.201200147}.

\bibitem[{\citenamefont{Dirac}(1958)}]{DiracHamGR}
\bibinfo{author}{\bibfnamefont{P.~A.~M.} \bibnamefont{Dirac}},
  \bibinfo{journal}{Proc.\ Roy.\ Soc.\ A} \textbf{\bibinfo{volume}{246}},
  \bibinfo{pages}{333} (\bibinfo{year}{1958}).

\bibitem[{\citenamefont{Katz}(1962)}]{Katz}
\bibinfo{author}{\bibfnamefont{J.}~\bibnamefont{Katz}},
  \bibinfo{journal}{Comptes Rendus Acad.\ Sci.\ Paris}
  \textbf{\bibinfo{volume}{254}}, \bibinfo{pages}{1386} (\bibinfo{year}{1962}).

\bibitem[{\citenamefont{Arnowitt et~al.}(1962)\citenamefont{Arnowitt, Deser,
  and Misner}}]{ADM}
\bibinfo{author}{\bibfnamefont{R.}~\bibnamefont{Arnowitt}},
  \bibinfo{author}{\bibfnamefont{S.}~\bibnamefont{Deser}}, \bibnamefont{and}
  \bibinfo{author}{\bibfnamefont{C.~W.} \bibnamefont{Misner}},
  \emph{\bibinfo{title}{The Dynamics of General Relativity}}
  (\bibinfo{publisher}{Wiley}, \bibinfo{address}{New York},
  \bibinfo{year}{1962}).

\bibitem[{\citenamefont{Green et~al.}(2012)\citenamefont{Green, Schwarz, and
  Witten}}]{GreenSchwarzWitten2012}
\bibinfo{author}{\bibfnamefont{M.~B.} \bibnamefont{Green}},
  \bibinfo{author}{\bibfnamefont{J.~H.} \bibnamefont{Schwarz}},
  \bibnamefont{and} \bibinfo{author}{\bibfnamefont{E.}~\bibnamefont{Witten}},
  \emph{\bibinfo{title}{Free bosonic strings}} (\bibinfo{publisher}{Cambridge
  University Press}, \bibinfo{year}{2012}), p. \bibinfo{pages}{57–120},
  Cambridge Monographs on Mathematical Physics.

\bibitem[{\citenamefont{Dirac}(1964)}]{dirac1964}
\bibinfo{author}{\bibfnamefont{P.}~\bibnamefont{Dirac}},
  \emph{\bibinfo{title}{Lectures on Quantum Mechanics}}, Belfer Graduate School
  of Science. Monographs series (\bibinfo{publisher}{Belfer Graduate School of
  Science, Yeshiva University}, \bibinfo{year}{1964}).

\bibitem[{\citenamefont{Bergmann}(1961)}]{BergmannTime}
\bibinfo{author}{\bibfnamefont{P.~G.} \bibnamefont{Bergmann}},
  \bibinfo{journal}{Rev.\ Mod.\ Phys.} \textbf{\bibinfo{volume}{33}},
  \bibinfo{pages}{510} (\bibinfo{year}{1961}).

\bibitem[{\citenamefont{Aharonov and Kaufherr}(1984)}]{QFR}
\bibinfo{author}{\bibfnamefont{Y.}~\bibnamefont{Aharonov}} \bibnamefont{and}
  \bibinfo{author}{\bibfnamefont{T.}~\bibnamefont{Kaufherr}},
  \bibinfo{journal}{Phys.\ Rev.\ D} \textbf{\bibinfo{volume}{30}},
  \bibinfo{pages}{368} (\bibinfo{year}{1984}).

\bibitem[{\citenamefont{Rovelli}(1991)}]{GeomObs2}
\bibinfo{author}{\bibfnamefont{C.}~\bibnamefont{Rovelli}},
  \bibinfo{journal}{Class.\ Quantum Grav.} \textbf{\bibinfo{volume}{8}},
  \bibinfo{pages}{317} (\bibinfo{year}{1991}).

\bibitem[{\citenamefont{Bartlett et~al.}(2007)\citenamefont{Bartlett, Rudolph,
  and Spekkens}}]{ReferenceFrames}
\bibinfo{author}{\bibfnamefont{S.~D.} \bibnamefont{Bartlett}},
  \bibinfo{author}{\bibfnamefont{T.}~\bibnamefont{Rudolph}}, \bibnamefont{and}
  \bibinfo{author}{\bibfnamefont{R.~W.} \bibnamefont{Spekkens}},
  \bibinfo{journal}{Rev.\ Mod.\ Phys.} \textbf{\bibinfo{volume}{79}},
  \bibinfo{pages}{555} (\bibinfo{year}{2007}), \eprint{arXiv:quant-ph/0610030}.

\bibitem[{\citenamefont{Giacomini et~al.}(2019)\citenamefont{Giacomini,
  Castro-Ruiz, and Brukner}}]{giacomini2019}
\bibinfo{author}{\bibfnamefont{F.}~\bibnamefont{Giacomini}},
  \bibinfo{author}{\bibfnamefont{E.}~\bibnamefont{Castro-Ruiz}},
  \bibnamefont{and}
  \bibinfo{author}{\bibfnamefont{{\v{C}}.}~\bibnamefont{Brukner}},
  \bibinfo{journal}{Nature Communications} \textbf{\bibinfo{volume}{10}},
  \bibinfo{pages}{494} (\bibinfo{year}{2019}), ISSN \bibinfo{issn}{2041-1723},
  \urlprefix\url{https://doi.org/10.1038/s41467-018-08155-0}.

\bibitem[{\citenamefont{Page and Wootters}(1983)}]{PageWootters}
\bibinfo{author}{\bibfnamefont{D.~N.} \bibnamefont{Page}} \bibnamefont{and}
  \bibinfo{author}{\bibfnamefont{W.~K.} \bibnamefont{Wootters}},
  \bibinfo{journal}{Phys.\ Rev.\ D} \textbf{\bibinfo{volume}{27}},
  \bibinfo{pages}{2885} (\bibinfo{year}{1983}).

\bibitem[{\citenamefont{Smith and Ahmadi}(2020)}]{smith2020}
\bibinfo{author}{\bibfnamefont{A.~R.~H.} \bibnamefont{Smith}} \bibnamefont{and}
  \bibinfo{author}{\bibfnamefont{M.}~\bibnamefont{Ahmadi}},
  \bibinfo{journal}{Nature Communications} \textbf{\bibinfo{volume}{11}},
  \bibinfo{pages}{5360} (\bibinfo{year}{2020}), ISSN \bibinfo{issn}{2041-1723},
  \urlprefix\url{https://doi.org/10.1038/s41467-020-18264-4}.

\bibitem[{\citenamefont{Strocchi}(1966)}]{strocchi1966}
\bibinfo{author}{\bibfnamefont{F.}~\bibnamefont{Strocchi}},
  \bibinfo{journal}{Rev.\ Mod.\ Phys.} \textbf{\bibinfo{volume}{38}},
  \bibinfo{pages}{36} (\bibinfo{year}{1966}).

\bibitem[{\citenamefont{Kibble}(1979)}]{kibble1979}
\bibinfo{author}{\bibfnamefont{T.~W.~B.} \bibnamefont{Kibble}},
  \bibinfo{journal}{Communications in Mathematical Physics}
  \textbf{\bibinfo{volume}{65}}, \bibinfo{pages}{189 } (\bibinfo{year}{1979}).

\bibitem[{\citenamefont{Ashtekar and Schilling}(1999)}]{ashtekar1999}
\bibinfo{author}{\bibfnamefont{A.}~\bibnamefont{Ashtekar}} \bibnamefont{and}
  \bibinfo{author}{\bibfnamefont{T.~A.} \bibnamefont{Schilling}},
  \emph{\bibinfo{title}{Geometrical Formulation of Quantum Mechanics}}
  (\bibinfo{publisher}{Springer New York}, \bibinfo{address}{New York, NY},
  \bibinfo{year}{1999}), pp. \bibinfo{pages}{23--65}, ISBN
  \bibinfo{isbn}{978-1-4612-1422-9},
  \urlprefix\url{https://doi.org/10.1007/978-1-4612-1422-9_3}.

\bibitem[{\citenamefont{Bojowald and Skirzewski}(2006)}]{bojowald2006}
\bibinfo{author}{\bibfnamefont{M.}~\bibnamefont{Bojowald}} \bibnamefont{and}
  \bibinfo{author}{\bibfnamefont{A.}~\bibnamefont{Skirzewski}},
  \bibinfo{journal}{Rev. Math. Phys.} \textbf{\bibinfo{volume}{18}},
  \bibinfo{pages}{713} (\bibinfo{year}{2006}).

\bibitem[{\citenamefont{Bojowald et~al.}(2009)\citenamefont{Bojowald,
  Sandh\"ofer, Skirzewski, and Tsobanjan}}]{EffCons}
\bibinfo{author}{\bibfnamefont{M.}~\bibnamefont{Bojowald}},
  \bibinfo{author}{\bibfnamefont{B.}~\bibnamefont{Sandh\"ofer}},
  \bibinfo{author}{\bibfnamefont{A.}~\bibnamefont{Skirzewski}},
  \bibnamefont{and}
  \bibinfo{author}{\bibfnamefont{A.}~\bibnamefont{Tsobanjan}},
  \bibinfo{journal}{Rev.\ Math.\ Phys.} \textbf{\bibinfo{volume}{21}},
  \bibinfo{pages}{111} (\bibinfo{year}{2009}), \eprint{arXiv:0804.3365}.

\bibitem[{\citenamefont{Bojowald and Tsobanjan}(2009)}]{EffConsRel}
\bibinfo{author}{\bibfnamefont{M.}~\bibnamefont{Bojowald}} \bibnamefont{and}
  \bibinfo{author}{\bibfnamefont{A.}~\bibnamefont{Tsobanjan}},
  \bibinfo{journal}{Phys.\ Rev.\ D} \textbf{\bibinfo{volume}{80}},
  \bibinfo{pages}{125008} (\bibinfo{year}{2009}), \eprint{arXiv:0906.1772}.

\bibitem[{\citenamefont{Gavrilov and
  Gitman}(2000{\natexlab{a}})}]{gavrilov2000a}
\bibinfo{author}{\bibfnamefont{S.~P.} \bibnamefont{Gavrilov}} \bibnamefont{and}
  \bibinfo{author}{\bibfnamefont{D.~M.} \bibnamefont{Gitman}},
  \bibinfo{journal}{International Journal of Modern Physics A}
  \textbf{\bibinfo{volume}{15}}, \bibinfo{pages}{4499}
  (\bibinfo{year}{2000}{\natexlab{a}}),
  \eprint{https://doi.org/10.1142/S0217751X00002378},
  \urlprefix\url{https://doi.org/10.1142/S0217751X00002378}.

\bibitem[{\citenamefont{Balsells and Bojowald}(2023)}]{balsells2023}
\bibinfo{author}{\bibfnamefont{J.}~\bibnamefont{Balsells}} \bibnamefont{and}
  \bibinfo{author}{\bibfnamefont{M.}~\bibnamefont{Bojowald}},
  \bibinfo{journal}{Phys. Rev. D} \textbf{\bibinfo{volume}{108}},
  \bibinfo{pages}{084030} (\bibinfo{year}{2023}),
  \urlprefix\url{https://link.aps.org/doi/10.1103/PhysRevD.108.084030}.

\bibitem[{\citenamefont{Hohmann et~al.}()\citenamefont{Hohmann, Pfeifer, and
  Wagner}}]{WeakEquivDispersion}
\bibinfo{author}{\bibfnamefont{M.}~\bibnamefont{Hohmann}},
  \bibinfo{author}{\bibfnamefont{C.}~\bibnamefont{Pfeifer}}, \bibnamefont{and}
  \bibinfo{author}{\bibfnamefont{F.}~\bibnamefont{Wagner}},
  \eprint{arXiv:2404.18811}.

\bibitem[{\citenamefont{Jackiw and Kerman}(1979)}]{VariationalEffAc}
\bibinfo{author}{\bibfnamefont{R.}~\bibnamefont{Jackiw}} \bibnamefont{and}
  \bibinfo{author}{\bibfnamefont{A.}~\bibnamefont{Kerman}},
  \bibinfo{journal}{Physics Letters A} \textbf{\bibinfo{volume}{71}},
  \bibinfo{pages}{158} (\bibinfo{year}{1979}), ISSN \bibinfo{issn}{0375-9601},
  \urlprefix\url{https://www.sciencedirect.com/science/article/pii/0375960179901518}.

\bibitem[{\citenamefont{Arickx et~al.}(1986)\citenamefont{Arickx, Broeckhove,
  Coene, and Van~Leuven}}]{GaussianDyn}
\bibinfo{author}{\bibfnamefont{F.}~\bibnamefont{Arickx}},
  \bibinfo{author}{\bibfnamefont{J.}~\bibnamefont{Broeckhove}},
  \bibinfo{author}{\bibfnamefont{W.}~\bibnamefont{Coene}}, \bibnamefont{and}
  \bibinfo{author}{\bibfnamefont{P.}~\bibnamefont{Van~Leuven}},
  \bibinfo{journal}{International Journal of Quantum Chemistry}
  \textbf{\bibinfo{volume}{30}}, \bibinfo{pages}{471} (\bibinfo{year}{1986}),
  \eprint{https://onlinelibrary.wiley.com/doi/pdf/10.1002/qua.560300741},
  \urlprefix\url{https://onlinelibrary.wiley.com/doi/abs/10.1002/qua.560300741}.

\bibitem[{\citenamefont{Prezhdo}(2006)}]{QHDTunneling}
\bibinfo{author}{\bibfnamefont{O.~V.} \bibnamefont{Prezhdo}},
  \bibinfo{journal}{Theoretical Chemistry Accounts}
  \textbf{\bibinfo{volume}{116}}, \bibinfo{pages}{206} (\bibinfo{year}{2006}),
  ISSN \bibinfo{issn}{1432-2234},
  \urlprefix\url{https://doi.org/10.1007/s00214-005-0032-x}.

\bibitem[{\citenamefont{Bayta\c{s} et~al.}(2020)\citenamefont{Bayta\c{s},
  Bojowald, and Crowe}}]{Bosonize}
\bibinfo{author}{\bibfnamefont{B.}~\bibnamefont{Bayta\c{s}}},
  \bibinfo{author}{\bibfnamefont{M.}~\bibnamefont{Bojowald}}, \bibnamefont{and}
  \bibinfo{author}{\bibfnamefont{S.}~\bibnamefont{Crowe}},
  \bibinfo{journal}{Annals of Physics} \textbf{\bibinfo{volume}{420}},
  \bibinfo{pages}{168247} (\bibinfo{year}{2020}), ISSN
  \bibinfo{issn}{0003-4916}, \eprint{arXiv:1810.12127},
  \urlprefix\url{https://www.sciencedirect.com/science/article/pii/S0003491620301810}.

\bibitem[{\citenamefont{Bayta\ifmmode~\mbox{\c{s}}\else \c{s}\fi{}
  et~al.}(2019)\citenamefont{Bayta\ifmmode~\mbox{\c{s}}\else \c{s}\fi{},
  Bojowald, and Crowe}}]{EffPotRealize}
\bibinfo{author}{\bibfnamefont{B.}~\bibnamefont{Bayta\ifmmode~\mbox{\c{s}}\else
  \c{s}\fi{}}}, \bibinfo{author}{\bibfnamefont{M.}~\bibnamefont{Bojowald}},
  \bibnamefont{and} \bibinfo{author}{\bibfnamefont{S.}~\bibnamefont{Crowe}},
  \bibinfo{journal}{Phys. Rev. A} \textbf{\bibinfo{volume}{99}},
  \bibinfo{pages}{042114} (\bibinfo{year}{2019}), \eprint{arXiv:1811.00505},
  \urlprefix\url{https://link.aps.org/doi/10.1103/PhysRevA.99.042114}.

\bibitem[{\citenamefont{Brizuela}(2014)}]{brizuela2014}
\bibinfo{author}{\bibfnamefont{D.}~\bibnamefont{Brizuela}},
  \bibinfo{journal}{Phys. Rev. D} \textbf{\bibinfo{volume}{90}},
  \bibinfo{pages}{125018} (\bibinfo{year}{2014}),
  \urlprefix\url{https://link.aps.org/doi/10.1103/PhysRevD.90.125018}.

\bibitem[{\citenamefont{Bojowald}(2022)}]{Description}
\bibinfo{author}{\bibfnamefont{M.}~\bibnamefont{Bojowald}},
  \bibinfo{journal}{J.\ Phys.\ A: Math.\ Theor.} \textbf{\bibinfo{volume}{55}},
  \bibinfo{pages}{504006} (\bibinfo{year}{2022}), \eprint{arXiv:2301.05138}.

\bibitem[{\citenamefont{Bojowald and Tsobanjan}(2021)}]{AlgebraicTime}
\bibinfo{author}{\bibfnamefont{M.}~\bibnamefont{Bojowald}} \bibnamefont{and}
  \bibinfo{author}{\bibfnamefont{A.}~\bibnamefont{Tsobanjan}},
  \bibinfo{journal}{Commun.\ Math.\ Phys.} \textbf{\bibinfo{volume}{382}},
  \bibinfo{pages}{547} (\bibinfo{year}{2021}), \eprint{arXiv:1906.04792}.

\bibitem[{\citenamefont{Bojowald and
  Tsobanjan}(2023{\natexlab{a}})}]{AlgebraicFrozen}
\bibinfo{author}{\bibfnamefont{M.}~\bibnamefont{Bojowald}} \bibnamefont{and}
  \bibinfo{author}{\bibfnamefont{A.}~\bibnamefont{Tsobanjan}},
  \bibinfo{journal}{Phys.\ Rev.\ D} \textbf{\bibinfo{volume}{107}},
  \bibinfo{pages}{024003} (\bibinfo{year}{2023}{\natexlab{a}}),
  \eprint{arXiv:2212.13961}.

\bibitem[{\citenamefont{Bojowald and
  Tsobanjan}(2023{\natexlab{b}})}]{TimeFluct}
\bibinfo{author}{\bibfnamefont{M.}~\bibnamefont{Bojowald}} \bibnamefont{and}
  \bibinfo{author}{\bibfnamefont{A.}~\bibnamefont{Tsobanjan}},
  \bibinfo{journal}{Quantum Rep.} \textbf{\bibinfo{volume}{5}},
  \bibinfo{pages}{22} (\bibinfo{year}{2023}{\natexlab{b}}),
  \eprint{arXiv:2211.04520}.

\bibitem[{\citenamefont{Smith and Ahmadi}(2019)}]{ClockSystemInt}
\bibinfo{author}{\bibfnamefont{A.~R.~H.} \bibnamefont{Smith}} \bibnamefont{and}
  \bibinfo{author}{\bibfnamefont{M.}~\bibnamefont{Ahmadi}},
  \bibinfo{journal}{Quantum} \textbf{\bibinfo{volume}{3}}, \bibinfo{pages}{160}
  (\bibinfo{year}{2019}), \eprint{arXiv:1712.00081}.

\bibitem[{\citenamefont{Lobo and Pfeifer}()}]{MuonDilation}
\bibinfo{author}{\bibfnamefont{I.~P.} \bibnamefont{Lobo}} \bibnamefont{and}
  \bibinfo{author}{\bibfnamefont{C.}~\bibnamefont{Pfeifer}},
  \eprint{arXiv:2406.05150}.

\bibitem[{\citenamefont{Gavrilov and
  Gitman}(2000{\natexlab{b}})}]{gavrilov2000b}
\bibinfo{author}{\bibfnamefont{S.~P.} \bibnamefont{Gavrilov}} \bibnamefont{and}
  \bibinfo{author}{\bibfnamefont{D.~M.} \bibnamefont{Gitman}},
  \bibinfo{journal}{Class. Quant. Grav.} \textbf{\bibinfo{volume}{17}},
  \bibinfo{pages}{L133} (\bibinfo{year}{2000}{\natexlab{b}}),
  \eprint{hep-th/0005249}.

\bibitem[{\citenamefont{Gitman and Tyutin}(1990)}]{gitman1990}
\bibinfo{author}{\bibfnamefont{D.~M.} \bibnamefont{Gitman}} \bibnamefont{and}
  \bibinfo{author}{\bibfnamefont{I.~V.} \bibnamefont{Tyutin}},
  \emph{\bibinfo{title}{Quantization of Fields with Constraints}}, Springer
  Series in Nuclear and Particle Physics (\bibinfo{publisher}{Springer Berlin,
  Heidelberg}, \bibinfo{year}{1990}), ISBN \bibinfo{isbn}{978-3-642-83938-2},
  \urlprefix\url{https://api.semanticscholar.org/CorpusID:115778106}.

\bibitem[{\citenamefont{Brown}(2022)}]{brown2022}
\bibinfo{author}{\bibfnamefont{J.~D.} \bibnamefont{Brown}},
  \bibinfo{journal}{Universe} \textbf{\bibinfo{volume}{8}}
  (\bibinfo{year}{2022}), ISSN \bibinfo{issn}{2218-1997},
  \urlprefix\url{https://www.mdpi.com/2218-1997/8/3/171}.

\bibitem[{\citenamefont{Hartle and Marolf}(1997)}]{GenRepIn}
\bibinfo{author}{\bibfnamefont{J.~B.} \bibnamefont{Hartle}} \bibnamefont{and}
  \bibinfo{author}{\bibfnamefont{D.}~\bibnamefont{Marolf}},
  \bibinfo{journal}{Phys.\ Rev.\ D} \textbf{\bibinfo{volume}{56}},
  \bibinfo{pages}{6247} (\bibinfo{year}{1997}), \eprint{gr-qc/9703021}.

\bibitem[{\citenamefont{Bjelakovi{\'{c}} and Stulpe}(2005)}]{bjelakovic2005}
\bibinfo{author}{\bibfnamefont{I.}~\bibnamefont{Bjelakovi{\'{c}}}}
  \bibnamefont{and} \bibinfo{author}{\bibfnamefont{W.}~\bibnamefont{Stulpe}},
  \bibinfo{journal}{International Journal of Theoretical Physics}
  \textbf{\bibinfo{volume}{44}}, \bibinfo{pages}{2041} (\bibinfo{year}{2005}),
  ISSN \bibinfo{issn}{1572-9575},
  \urlprefix\url{https://doi.org/10.1007/s10773-005-8982-2}.

\end{thebibliography}

\end{document}